\documentclass[superscriptaddress, aps, prx, longbibliography]{revtex4-2}

\usepackage{graphicx}
\usepackage{physics}
\usepackage{amsmath}
\usepackage{array}
\usepackage{siunitx}
\usepackage[section]{placeins}
\usepackage{hyperref}

\usepackage{natbib}

\begin{document}

\title{Characterisation of a levitated sub-mg ferromagnetic cube \\in a planar alternating-current magnetic Paul trap}

\date{August 2024}

\author{Martijn Janse}
\author{Eli van der Bent}
\author{Mart Laurman}
\author{Robert Smit}
\author{Bas Hensen} \email{hensen@physics.leidenuniv.nl}
\address{Leiden Institute of Physics, Leiden University, P.O. Box 9504, 2300 RA Leiden, The Netherlands}

\begin{abstract}
Microscopic levitated objects are a promising platform for inertial sensing, testing gravity at small scales, optomechanics in the quantum regime, and large-mass superpositions. However, existing levitation techniques harnessing optical and electrical fields suffer from noise induced by elevated internal temperatures and charge noise, respectively. Meissner-based magnetic levitation circumvents both sources of decoherence but requires cryogenic environments. Here we characterize a sub-mg ferromagnetic cube levitated in an alternating-current planar magnetic Paul trap at room temperature. We show behavior in line with the Mathieu equations and quality factors of up to 2500 for the librational modes. Besides technological sensing applications, this technique sets out a path for MHz librational modes in the micron-sized particle limit, allowing for magnetic coupling to superconducting circuits and spin-based quantum systems. \end{abstract}

\maketitle

Over the past decades, levitated mechanical resonators have propelled both technological and scientific developments \cite{gonzalez-ballestero_levitodynamics_2021}. They have enabled high-precision sensing applications \cite{wang_mechanical_2024, ahn_ultrasensitive_2020, moore_search_2014, hempston_force_2017, ricci_accurate_2019, slezak_cooling_2018, monteiro_optical_2017} and are a promising platform for quantum matter-wave interferometry proposals \cite{romero-isart_large_2011, kaltenbaek_macroscopic_2012, bateman_near-field_2014, wan_free_2016,pino_-chip_2018}. More recently, levitated resonators have increasingly come into play as a candidate platform to study the interface between quantum mechanics and general relativity \cite{nimmrichter_macroscopicity_2013, bassi_models_2013, bose_spin_2017}. \\

Optical tweezers are one of the platforms used to levitate objects ranging from single atoms to microspheres \cite{ashkin_acceleration_1970, chu_experimental_1986, chu_nobel_1998, phillips_nobel_1998, hu_creation_2017}, advancing research into dark matter detection \cite{moore_searching_2021}, atomic clocks \cite{ludlow_optical_2015} and the quantum limit of macroscopic masses \cite{gieseler_subkelvin_2012, gao_feedback_2024, tebbenjohanns_quantum_2021, piotrowski_simultaneous_2023}. However, larger objects have increased internal temperatures caused by absorption of the laser light, which limits the type of objects that can be levitated \cite{rahman_burning_2016}, and causes additional noise due to black-body radiation \cite{romero-isart_large_2011}. This can be circumvented by levitating charged particles in an electrical Paul trap \cite{paul_electromagnetic_1990}, which has applications in spin cooling of motion, trapping electrons and quantum computing \cite{ostermayr_transportable_2018, dania_optical_2021, jain_penning_2024, delord_spin-cooling_2020, alda_trapping_2016, matthiesen_trapping_2021}. Still, for future large-mass quantum coherent experiments, charge-neutral particles may be crucial.\\

Therefore, magnetic levitation has recently gained interest \cite{hofer_high-_2023, gutierrez_latorre_superconducting_2023,fuchs_measuring_2024, timberlake_acceleration_2019, timberlake_linear_2023, vinante_ultralow_2020, gieseler_single-spin_2020}. Diamagnetic objects can be levitated magnetically using static magnet fields. At room temperature this typically leads to weak confinement \cite{hsu_cooling_2016} but at cryogenic temperatures, the Meissner effect allows for stronger confinement of superconducting levitated particles \cite{hofer_high-_2023, gutierrez_latorre_superconducting_2023}, or vice versa of ferromagnetic particles above a superconducting substrate \cite{fuchs_measuring_2024, timberlake_acceleration_2019, timberlake_linear_2023, vinante_ultralow_2020, gieseler_single-spin_2020}.\\

Here, we investigate levitation of ferromagnetic particles using alternating magnetic fields. Following up on several setups,
levitating ferromagnetic particles with extra non-magnetic confining forces \cite{barry_magneto-optical_2014}, rotating ferromagnets \cite{hermansen_magnetic_2023, sakuma_three-dimensional_2023, simon_spin_1997} or alternating-current (AC) magnetic trapping fields \cite{sackett_magnetic_1993, cornell_multiply_1991}, Perdriat et al. \cite{perdriat_planar_2023} have recently put forward a proposal for an on-chip planar AC magnetic Paul trap (MPT). Besides harnessing the advantages of magnetic levitation at room temperature, the planar AC MPT also allows for control over a wide range of trap properties. Moreover, the platform can be miniaturized, which may allow for magnetic-field tunable librational degrees of freedom with eigenfrequencies of up to MHz.\\

In this work, we make a first step towards an on-chip implementation by demonstrating and characterizing the planar AC MPT at intermediate scale. We levitate a 250 $\mu$m hard ferromagnet in a trap on a printed circuit board (PCB) with millimeter-sized current loops.\\

Since Earnshaw's theorem states that a ferromagnetic particle cannot be trapped by a static magnetic field in free space, we employ AC magnetic fields to generate a ponderomotive restoring force on the particle. The planar AC MPT in this work uses two magnetic fields. First, we use an AC magnetic field $\vec{B_1}$, generated by an inner loop carrying an alternating trap current $I_{\text{trap}}(\Omega)$, and an outer loop carrying $-\xi I_{\text{trap}}(\Omega)$ with $\Omega/(2\pi)$ the trap frequency and $\xi$ the ratio between the currents. Lastly, to stabilize the angular direction of the levitated magnet and counteract the gravitational force, we position two coils in Helmholtz configuration to generate a magnetic field in the z-direction, with $\vec{B}_{0} = (B_{0} + B_{0}^{'}z)\hat{z}$. The gradient between the Helmholtz coils allows for tuning of the levitation height. For $|\vec{B_0}| \gg |\vec{B_1}|$, we expect stable levitation in the center of the trap.\\

\begin{figure}
    \centering
    \includegraphics[scale=1.0]{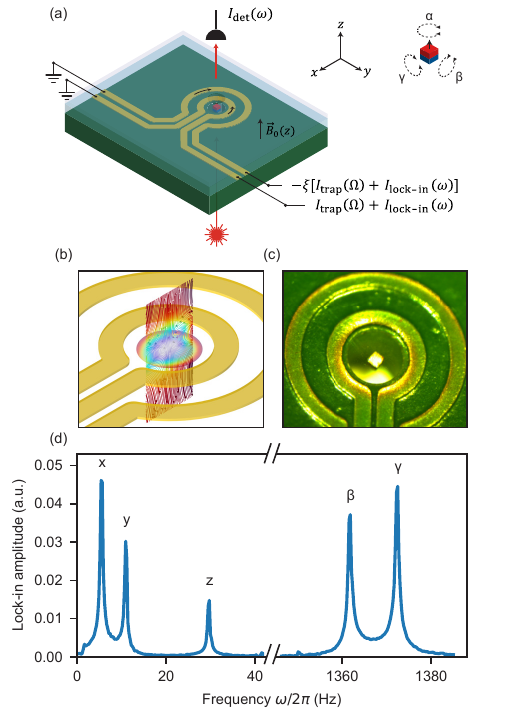}
    \caption{\textbf{(a)} Schematic overview of the set-up. A ferromagnetic cube is levitated by two alternating currents running in opposite directions through gold tracks on a printed circuit board (PCB, dark green). The trapping volume is formed by a hole in the PCB and a top cover glass (light blue), and capped above and below by two additional thin glass slides (light gray). A photodiode detects its motion via a laser illuminating the magnet from below. \textbf{(b)} Finite-element simulation of the magnetic field norm and field lines generated by the current-carrying loops for current ratio $\xi = 2.35$. \textbf{(c)} Photo of a levitating ferromagnetic cube in the trap. \textbf{(d)} Motional spectrum showing three vibrational and two librational modes.}
    \label{Figure1}
\end{figure}

From the Mathieu equations we obtain the following expressions for the center-of-mass eigenmodes or secular frequencies $\omega_\text{i}/(2\pi)$ of the vibrational modes in the x-, y- and z-direction \cite{sackett_magnetic_1993,perdriat_planar_2023}:

\begin{equation}
\label{vib_modes}
    \omega_\text{z} = 2\omega_{\text{x,y}} = \frac{\Omega}{2}\frac{|q_\text{z}|}{\sqrt{2}} = \frac{|B''_1|B_{\text{sat}}}{\mu_0\rho_m\Omega\sqrt{2}},
\end{equation}

and those of the librational modes in the $\beta$ and $\gamma$ direction (see Fig.~\ref{Figure1}(a)) following the definitions given in \cite{perdriat_planar_2023}:

\begin{equation}
\label{lib_modes}
    \omega_{\beta} = \omega_{\gamma} = \sqrt{\frac{5}{2}\frac{B_0B_{\text{sat}}}{\mu_0\rho_m a^2}},
\end{equation}

where $q_\text{z} = -2q_\text{x,y} = 2B^{''}_1 B_\text{sat}/(\mu_0 \rho_m \Omega^2)$, $B^{''}_1$ is the magnetic field curvature of $\vec{B}_1$, $B_{\text{sat}}$ the magnetization field of the trapped magnet, $\mu_0$ the magnetic permeability of vacuum, $
\rho_m$ the density of the magnet, $\Omega/(2\pi)$ the trap frequency and $a$ the radius of a magnetic sphere. Due to the symmetry of the MPT, we cannot derive an expression for the librational $\alpha$ eigenmode, because the torque in the $\alpha$-direction $\Gamma_\alpha = 0$ for a point particle. For this derivation, we assume to obey the stability criterion following from the Mathieu equations, such that $q_\text{z} < 0.4$ \cite{dehmelt_radiofrequency_1968, perdriat_planar_2023}.\\

In Fig.~\ref{Figure1}(a)-(c) we show the experimental realization of the proposed AC planar MPT at intermediate scale. The two current loops are made of copper tracks of width 0.3 mm and height 0.035 mm on a PCB made of FR4. The currents generate a magnetic field $\vec{B_1}$ which is simulated via a finite-element method in Fig.~\ref{Figure1}(b). The inner respectively outer radius $R_1 = \frac{R_2}{2}$ = 0.7 mm is measured from the center of the trap to the center of the track. A voltage from a function generator is amplified to high current via an AC power amplifier and split up via a variable resistor into two parallel currents, $I_{\text{trap}}$ and $-\xi I_{\text{trap}}$. The Helmholtz coils with radius $R_{\text{HHC}} = 10$ mm, consisting of $N=835$ copper-wire windings, are placed above and below the PCB and are expected to generate a field $B_{0}$ of 1-10 mT around the center of the MPT. The magnetic fields $B_0$ and $B_1$ reported in this paper are calculated values (see supplementary material, Fig. S3).\\

In the MPT we levitate a nickel-coated cubic Nd$_2$Fe$_{14}$B magnet with edge lengths of $a = 250$ $\mu m$ and magnetization field $B_\text{sat} \sim 1.4$T. We illuminate the levitated magnetic with a HeNe laser ($\lambda = 632.8$ nm), and measure the diffracted light on a photodiode. The photodiode is connected as input to a lock-in amplifier, with a high-pass filter at cut-off frequency of $300$ Hz for measuring the librational modes. The output $I_\text{lock-in}(\omega)$ is connected in parallel to the function generator output. \\

In the left part of Fig.~\ref{Figure1}(d), we reveal three vibrational modes at $(\omega_\text{x}, \omega_\text{y}, \omega_\text{z}) = 2\pi\cdot(5.48, 10.9, 29.6)$ Hz measured at $\Omega/(2\pi) = 120$ Hz, $I_\text{trap} = 1.07$ A, $B_0 = 5.6$ mT and $B_{0}^{'} = 81$ mT/m, and two librational modes at $(\omega_{\beta}, \omega_{\gamma}) = 2\pi\cdot(1362, 1372)$ Hz, measured at $B_0 = 9.4$ mT.\\

\begin{figure}
    \centering
    \includegraphics[scale=1.4]{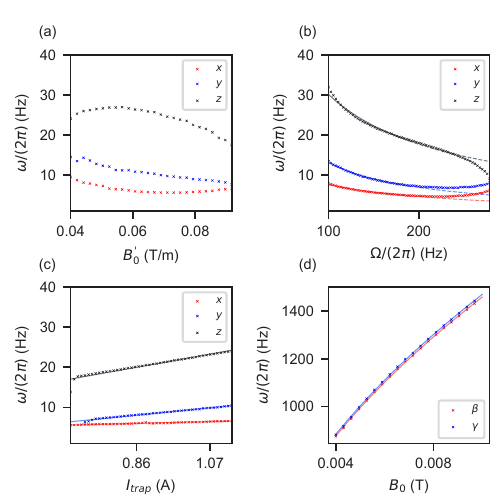}
    \caption{Dependence of eigenmode frequencies on trap parameters. Full spectral data is available in the supplementary material, Fig. S1 and S2. Magnetic field gradient $B_{0}^{'}$, magnetic field strength $B_{0}$ and trap current $I_{\text{trap}}$ are calibrated according to Fig. S3 in the supplementary material. All measurements are performed at current ratio $\xi = 2.2 \pm 0.05$. For \textbf{(b)-(d)} we set $B_{0}^{'} = 81$ mT/m. For \textbf{(a)} and \textbf{(c)} we set the trap frequency at $\Omega/(2\pi) = 150$ Hz, while for \textbf{(d)} we set $\Omega/(2\pi) = 140$ Hz. For \textbf{(a)}, \textbf{(b)} and \textbf{(d)} we set $I_{\text{trap}} = 0.97$ A. For \textbf{(a)-(c)} we set $B_{0} = 5.6$ mT. We use Eqns. \ref{vib_modes} and \ref{lib_modes} to fit \textbf{(b)} an inverse function for $100$ Hz $<$ $\Omega/(2\pi) < 200$ Hz, \textbf{(c)} a linear function and \textbf{(d)} a square root function.}
    \label{Figure2}
\end{figure}

Due to a slight asymmetry in the trap current loops, we also see confinement in the $\alpha$-direction, leading to a weakly-coupled librational mode at $\omega_{\alpha}/(2\pi) \sim 1-3$ Hz. It can be detected via mode coupling to the librational modes in the $\beta$- and $\gamma$-direction, which is verified by camera-based detection of the magnet motion in the $\alpha$-direction.\\

We check whether the magnet motion behaves according to the Mathieu equations by experimentally testing the relations in Eqns. \ref{vib_modes} and $\ref{lib_modes}$. In Fig.~\ref{Figure2}(a) the dependency of the vibrational eigenmode frequencies on the magnetic field gradient $B_{0}^{'}$ is shown. By changing $B_{0}^{'}$, we can alter the levitation height of the magnet in the MPT, such that it experiences different magnetic forces from the AC magnetic fields. From Eqn. $\ref{vib_modes}$ we know that $\omega_\text{x,y,z} \sim |B_1''|$, but the dependency of $|B_1^{''}|$ on $B_{0}^{'}$ and thus the levitation height is not trivial.\\

We observe that $\omega_\text{z}$ reaches a maximum at $B_{0}^{'} = 53$ mT/m, whereas $\omega_\text{x}$ is minimal for $B_{0}^{'} = 70$ mT/m. For the remaining eigenmode $\omega_\text{y}$ we cannot derive a minimum as the signal is too weak. Moreover, we see that the scaling of $\omega_\text{z}$ with $B_{0}^{'}$ is flipped in comparison to the scaling of $\omega_\text{x,y}$. At the maximum of $\omega_\text{z}$, the MPT is expected to be the most stiff in the z-direction for $B_{0}^{'} = 53$ mT/m, which may coincide with the height where the magnet levitates in the center of the MPT. However, we also see that at this levitation height the magnet levitates less stably than for $B_{0}^{'} = 81$ mT/m, as shown by the smaller ranges of trapping current $I_{\text{trap}}$ and trapping frequency $\Omega$ over which we can sweep (see supplementary material, Fig. S2).\\

Fig.~\ref{Figure2}(b) shows the dependence of the vibrational eigenmode frequencies on the trap frequency $\Omega$. From Eqn. \ref{vib_modes} we expect it to be inversely proportional to $\Omega$. This seems to be true for $\Omega/(2\pi) = 100$ Hz up to $\Omega/(2\pi) = 200$ Hz. For $\Omega/(2\pi) > 200$ Hz, the $\Omega$ dependency changes such that all three vibrational eigenmode frequencies appear to converge, at which point the magnet escapes the trap. A theoretical framework to explain this behavior is not provided by the current model and has yet to be devised. The slope ratio deviates from the theoretical value $\omega_\text{x}:\omega_\text{y}:\omega_\text{z} = 1:1:2$ (Eqn. \ref{vib_modes}), possibly because of trap asymmetry and the resulting out-of-center position of the magnet (see supplementary material, Fig. S4).\\

Fig.~\ref{Figure2}(c) shows the dependency of the vibrational eigenmode frequencies on the trap current $I_\text{trap}$. According to Eqn. $\ref{vib_modes}$, we expect a linear relationship, since $|B_1''| \sim I_\text{trap}$. We indeed see a linear relationship between the eigenmode frequencies and the trap current from $I_\text{trap} = 0.64$ A up to $I_\text{trap} = 1.13$ A. When $I_\text{trap}$ is lowered, the magnet again moves out of center, which may explain the deviation of the slopes from the theoretical ratio based on symmetry arguments, $\omega_\text{x}:\omega_\text{y}:\omega_\text{z} = 1:1:2$ (cf. Eqn. \ref{vib_modes}). Also, an asymmetry between $\omega_\text{x}$ and $\omega_\text{y}$ is expected because of the asymmetry due to the opening in the trap loops (see supplementary material, Fig. S4).\\

Via Eqn. \ref{vib_modes} we can calculate $|B_1^{''}|$ from the vibrational z-mode at $\omega_\text{z}/(2\pi) = 29.6$ Hz, giving $|B_1^{''}| = 1335$ T/m$^2$ for $\Omega/(2\pi) = 120$ Hz, $B_\text{sat} = 1.4$ T, $\mu_0 = 4\pi \cdot 10^{-7}$ N/A$^2$ and $\rho_m = 7.5 \cdot 10^3$ kg/m$^3$. This value for $|B_1^{''}|$ is measured at $I_\text{trap} = 1.07$ A. In the supplementary material, Fig. S3, we show that this value is within a factor 2 of the predicted value based on a COMSOL Multiphysics model of the planar AC MPT.\\

In Fig.~\ref{Figure2}(d) we examine the dependence of the librational eigenmode frequencies in the $\beta$- and $\gamma$-direction on the magnetic field $B_{0}$. According to Eqn. $\ref{lib_modes}$, we expect that $\omega_{\beta,\gamma} \sim \sqrt{B_{0}}$. We indeed find this relationship in the range $B_{0} = 3.8$ mT up to $B_{0} = 10.8$ mT. As expected from Eqn. $\ref{lib_modes}$, the magnitude and slope of $\beta$ and $\gamma$ are similar. The minimal $B_{0}$ value (at constant $B_{0}^{'}$) is the value where $|\vec{B_1}| \geq |\vec{B_0}|$, such that the angular dynamics of the magnetic dipole moment $\vec{\mu}$ becomes unstable. For our experiment, the maximal $B_{0}$ value is set by the heat dissipation in the Helmholtz coils.\\

\begin{figure}
    \centering
    \includegraphics[scale=1.5]{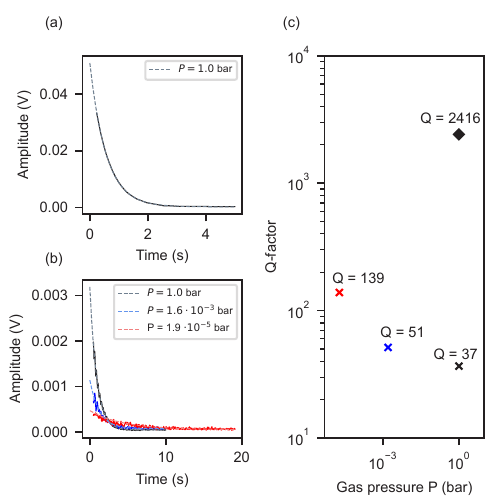}
    \caption{Vibrational and librational mode quality factors. \textbf{(a)} Averaged ringdown measurement signal of the librational $\gamma$-mode at $\Omega/(2\pi) = 1372.4$ Hz at gas pressure $P = 1.0$ bar fitted with an exponential decay function. \textbf{(b)} Averaged ringdown measurement signal of the vibrational y-mode at $\Omega/(2\pi) = 13.0$ Hz (for $P = 1.0$ bar and $P = 1.6 \cdot 10^{-3}$ bar) and at $\Omega/(2\pi) = 10.4$ Hz (for $P = 1.9 \cdot 10^{-5}$ bar) fitted with an exponential decay function. \textbf{(c)} Dependence of Q-factors on P for librational mode (diamond) and vibrational mode (crosses). The colors match those in \textbf{(a)-(b)}.}
    \label{Figure3}
\end{figure}

In Fig.~\ref{Figure3}(a)-(b) the averaged ringdown measurements of a librational and vibrational mode are shown, from which the average decay time $\bar{\tau}$ and quality factor $Q = \pi \omega_\text{i}/(2\pi)\bar{\tau}$ is determined. We observe that at ambient pressure the Q-factors of the librational eigenmodes are a factor 100 higher than the vibrational eigenmodes.\\

For the vibrational modes, the Q-factors are measured to be only weakly dependent on the gas pressure, as shown in Fig.~\ref{Figure3}(c). If the Q-factor of the levitated magnet were to be dominated by gas damping, we would expect that $Q \sim \gamma^{-1} \sim P^{-1}$ where $\gamma$ is the damping coefficient and P the gas pressure \cite{ranjit_zeptonewton_2016, schmole_micromechanical_2016}. Although the gas pressure decreases by five orders of magnitude, the Q-factor increases by less than one order of magnitude. Other dissipative channels therefore dominate, such as eddy currents generated by the ferromagnetic cube in the trap loops or eddy currents generated by the AC magnetic fields in the ferromagnetic cube. One would then also expect the Q-factor to saturate for lower gas pressure, which is not observed. This might be explained by the increased temperature of the copper tracks at low gas pressure, resulting in higher resistivity and thus lower eddy current damping.\\

The planar AC MPT characterized here provides a versatile highly tunable room-temperature magnetic levitation platform. In its present form, it serves as a platform to study the complex behaviour of magnetic dipole objects with six degrees of freedom. Due to its low cost of implementation (estimated cost of a few hundred euros), it may also serve as a setup for teaching purposes.\\ 

Presently, the Q-factors of the eigenmodes are prohibitively low, compared to other magnetic levitation platforms \cite{fuchs_measuring_2024, timberlake_linear_2023, gutierrez_latorre_superconducting_2023, hofer_high-_2023}, preventing access the quantum regime. This can be improved by reducing the eddy current damping, e.g. by removing the coating
of the magnet, using smaller or electrically insulating magnets or increasing the inner radius of the trapping coils. A major improvement would be the use of superconducting current tracks, at the cost of having to operate the trap at cryogenic temperatures.\\

The eigenmode frequencies can be increased by further miniaturization of the planar AC MPT. For the librational eigenmode frequencies, according to Eqn. \ref{lib_modes} we have $\omega_{\beta,\gamma} \sim \sqrt{B_{0}}/a$. If we further increase $B_{0}$ from 10 mT to 1 T and decrease the particle dimension from 250 $\mu$m to 1 $\mu$m, we can reach librational eigenmode frequencies up to 3 MHz.\\

These improvements may allow the planar AC MPT to be utilized in the future for coupling to spin-based quantum systems, such as NV centers \cite{huillery_spin_2020, delord_spin-cooling_2020} or, in cryogenic environments, optomechanical coupling to superconducting circuits. The relatively large magnetic dipole of the ferromagnet, as compared to the induced magnetic dipole of diamagnetically levitated superconducting particles, allows for a stronger coupling to superconducting quantum interference devices (SQUIDs) \cite{gutierrez_latorre_superconducting_2023, schmidt_remote_2024}. Therefore, the miniaturized on-chip MPT may be a valuable addition to the already existing levitation platforms to study fundamental physics.\\

See the supplementary material for the spectral information of the parameters sweeps, the calibration of $B_0$, $B_0^{'}$ and $|B_1^{''}|$ and the time-domain simulations of a ferromagnetic cube in a planar AC MPT.\\

We thank D.G. Uitenbroek and T.H. Oosterkamp for their contributions to the manuscript. This work was supported by the European Union (ERC StG, CLOSEtoQG, Project 101041115).

\section*{AUTHOR DECLARATIONS}
\subsection*{Conflict of interest}
The authors have no conflicts to disclose.

\subsection*{Author contributions}
\textbf{Martijn Janse}: Conceptualization (equal); Data curation  (equal); Formal analysis (equal); Investigation (equal); Methodology (equal); Project administration (lead); Software (equal); Validation (equal); Visualization (lead); Writing - Original Draft Preparation (lead); Writing - Review \& Editing (lead). \textbf{Eli van der Bent}: Conceptualization (equal); Data curation (equal); Formal analysis (equal); Investigation (equal); Methodology (equal); Software (equal); Visualization (supporting); Writing - Review \& Editing (supporting). \textbf{Mart Laurman}: Data curation (equal); Formal analysis (equal); Investigation (equal); Software (equal), Visualization (supporting); Writing - Review \& Editing (supporting). \textbf{Robert Smit}: Visualization (supporting); Writing - Review \& Editing (supporting) \textbf{Bas Hensen}: Conceptualization (equal); Data curation (equal); Formal analysis (equal); Funding acquisition (lead); Methodology (equal); Resources (lead); Supervision (lead); Validation (equal); Visualization (supporting); Writing - Original Draft Preparation (supporting); Writing - Review \& Editing (supporting).

\section*{DATA AVAILABILITY}
The data that support the findings of this study are available from the corresponding author upon reasonable request.

\section*{REFERENCES}
\bibliography{01_bare-references}

\begin{thebibliography}{55}%
\makeatletter
\providecommand \@ifxundefined [1]{%
 \@ifx{#1\undefined}
}%
\providecommand \@ifnum [1]{%
 \ifnum #1\expandafter \@firstoftwo
 \else \expandafter \@secondoftwo
 \fi
}%
\providecommand \@ifx [1]{%
 \ifx #1\expandafter \@firstoftwo
 \else \expandafter \@secondoftwo
 \fi
}%
\providecommand \natexlab [1]{#1}%
\providecommand \enquote  [1]{``#1''}%
\providecommand \bibnamefont  [1]{#1}%
\providecommand \bibfnamefont [1]{#1}%
\providecommand \citenamefont [1]{#1}%
\providecommand \href@noop [0]{\@secondoftwo}%
\providecommand \href [0]{\begingroup \@sanitize@url \@href}%
\providecommand \@href[1]{\@@startlink{#1}\@@href}%
\providecommand \@@href[1]{\endgroup#1\@@endlink}%
\providecommand \@sanitize@url [0]{\catcode `\\12\catcode `\$12\catcode `\&12\catcode `\#12\catcode `\^12\catcode `\_12\catcode `\%12\relax}%
\providecommand \@@startlink[1]{}%
\providecommand \@@endlink[0]{}%
\providecommand \url  [0]{\begingroup\@sanitize@url \@url }%
\providecommand \@url [1]{\endgroup\@href {#1}{\urlprefix }}%
\providecommand \urlprefix  [0]{URL }%
\providecommand \Eprint [0]{\href }%
\providecommand \doibase [0]{https://doi.org/}%
\providecommand \selectlanguage [0]{\@gobble}%
\providecommand \bibinfo  [0]{\@secondoftwo}%
\providecommand \bibfield  [0]{\@secondoftwo}%
\providecommand \translation [1]{[#1]}%
\providecommand \BibitemOpen [0]{}%
\providecommand \bibitemStop [0]{}%
\providecommand \bibitemNoStop [0]{.\EOS\space}%
\providecommand \EOS [0]{\spacefactor3000\relax}%
\providecommand \BibitemShut  [1]{\csname bibitem#1\endcsname}%
\let\auto@bib@innerbib\@empty
\bibitem [{\citenamefont {Gonzalez-Ballestero}\ \emph {et~al.}(2021)\citenamefont {Gonzalez-Ballestero}, \citenamefont {Aspelmeyer}, \citenamefont {Novotny}, \citenamefont {Quidant},\ and\ \citenamefont {Romero-Isart}}]{gonzalez-ballestero_levitodynamics_2021}%
  \BibitemOpen
  \bibfield  {author} {\bibinfo {author} {\bibfnamefont {C.}~\bibnamefont {Gonzalez-Ballestero}}, \bibinfo {author} {\bibfnamefont {M.}~\bibnamefont {Aspelmeyer}}, \bibinfo {author} {\bibfnamefont {L.}~\bibnamefont {Novotny}}, \bibinfo {author} {\bibfnamefont {R.}~\bibnamefont {Quidant}},\ and\ \bibinfo {author} {\bibfnamefont {O.}~\bibnamefont {Romero-Isart}},\ }\bibfield  {title} {\bibinfo {title} {Levitodynamics: {Levitation} and control of microscopic objects in vacuum},\ }\href {https://doi.org/10.1126/science.abg3027} {\bibfield  {journal} {\bibinfo  {journal} {Science}\ }\textbf {\bibinfo {volume} {374}},\ \bibinfo {pages} {eabg3027} (\bibinfo {year} {2021})}\BibitemShut {NoStop}%
\bibitem [{\citenamefont {Wang}\ \emph {et~al.}(2024)\citenamefont {Wang}, \citenamefont {Penny}, \citenamefont {Recoaro}, \citenamefont {Siegel}, \citenamefont {Tseng},\ and\ \citenamefont {Moore}}]{wang_mechanical_2024}%
  \BibitemOpen
  \bibfield  {author} {\bibinfo {author} {\bibfnamefont {J.}~\bibnamefont {Wang}}, \bibinfo {author} {\bibfnamefont {T.}~\bibnamefont {Penny}}, \bibinfo {author} {\bibfnamefont {J.}~\bibnamefont {Recoaro}}, \bibinfo {author} {\bibfnamefont {B.}~\bibnamefont {Siegel}}, \bibinfo {author} {\bibfnamefont {Y.-H.}\ \bibnamefont {Tseng}},\ and\ \bibinfo {author} {\bibfnamefont {D.~C.}\ \bibnamefont {Moore}},\ }\bibfield  {title} {\bibinfo {title} {Mechanical {Detection} of {Nuclear} {Decays}},\ }\href {https://doi.org/10.1103/PhysRevLett.133.023602} {\bibfield  {journal} {\bibinfo  {journal} {Physical Review Letters}\ }\textbf {\bibinfo {volume} {133}},\ \bibinfo {pages} {023602} (\bibinfo {year} {2024})}\BibitemShut {NoStop}%
\bibitem [{\citenamefont {Ahn}\ \emph {et~al.}(2020)\citenamefont {Ahn}, \citenamefont {Xu}, \citenamefont {Bang}, \citenamefont {Ju}, \citenamefont {Gao},\ and\ \citenamefont {Li}}]{ahn_ultrasensitive_2020}%
  \BibitemOpen
  \bibfield  {author} {\bibinfo {author} {\bibfnamefont {J.}~\bibnamefont {Ahn}}, \bibinfo {author} {\bibfnamefont {Z.}~\bibnamefont {Xu}}, \bibinfo {author} {\bibfnamefont {J.}~\bibnamefont {Bang}}, \bibinfo {author} {\bibfnamefont {P.}~\bibnamefont {Ju}}, \bibinfo {author} {\bibfnamefont {X.}~\bibnamefont {Gao}},\ and\ \bibinfo {author} {\bibfnamefont {T.}~\bibnamefont {Li}},\ }\bibfield  {title} {\bibinfo {title} {Ultrasensitive torque detection with an optically levitated nanorotor},\ }\href {https://doi.org/10.1038/s41565-019-0605-9} {\bibfield  {journal} {\bibinfo  {journal} {Nature Nanotechnology}\ }\textbf {\bibinfo {volume} {15}},\ \bibinfo {pages} {89} (\bibinfo {year} {2020})}\BibitemShut {NoStop}%
\bibitem [{\citenamefont {Moore}\ \emph {et~al.}(2014)\citenamefont {Moore}, \citenamefont {Rider},\ and\ \citenamefont {Gratta}}]{moore_search_2014}%
  \BibitemOpen
  \bibfield  {author} {\bibinfo {author} {\bibfnamefont {D.~C.}\ \bibnamefont {Moore}}, \bibinfo {author} {\bibfnamefont {A.~D.}\ \bibnamefont {Rider}},\ and\ \bibinfo {author} {\bibfnamefont {G.}~\bibnamefont {Gratta}},\ }\bibfield  {title} {\bibinfo {title} {Search for {Millicharged} {Particles} {Using} {Optically} {Levitated} {Microspheres}},\ }\href {https://doi.org/10.1103/PhysRevLett.113.251801} {\bibfield  {journal} {\bibinfo  {journal} {Physical Review Letters}\ }\textbf {\bibinfo {volume} {113}},\ \bibinfo {pages} {251801} (\bibinfo {year} {2014})}\BibitemShut {NoStop}%
\bibitem [{\citenamefont {Hempston}\ \emph {et~al.}(2017)\citenamefont {Hempston}, \citenamefont {Vovrosh}, \citenamefont {Toroš}, \citenamefont {Winstone}, \citenamefont {Rashid},\ and\ \citenamefont {Ulbricht}}]{hempston_force_2017}%
  \BibitemOpen
  \bibfield  {author} {\bibinfo {author} {\bibfnamefont {D.}~\bibnamefont {Hempston}}, \bibinfo {author} {\bibfnamefont {J.}~\bibnamefont {Vovrosh}}, \bibinfo {author} {\bibfnamefont {M.}~\bibnamefont {Toroš}}, \bibinfo {author} {\bibfnamefont {G.}~\bibnamefont {Winstone}}, \bibinfo {author} {\bibfnamefont {M.}~\bibnamefont {Rashid}},\ and\ \bibinfo {author} {\bibfnamefont {H.}~\bibnamefont {Ulbricht}},\ }\bibfield  {title} {\bibinfo {title} {Force sensing with an optically levitated charged nanoparticle},\ }\href {https://doi.org/10.1063/1.4993555} {\bibfield  {journal} {\bibinfo  {journal} {Applied Physics Letters}\ }\textbf {\bibinfo {volume} {111}},\ \bibinfo {pages} {133111} (\bibinfo {year} {2017})}\BibitemShut {NoStop}%
\bibitem [{\citenamefont {Ricci}\ \emph {et~al.}(2019)\citenamefont {Ricci}, \citenamefont {Cuairan}, \citenamefont {Conangla}, \citenamefont {Schell},\ and\ \citenamefont {Quidant}}]{ricci_accurate_2019}%
  \BibitemOpen
  \bibfield  {author} {\bibinfo {author} {\bibfnamefont {F.}~\bibnamefont {Ricci}}, \bibinfo {author} {\bibfnamefont {M.~T.}\ \bibnamefont {Cuairan}}, \bibinfo {author} {\bibfnamefont {G.~P.}\ \bibnamefont {Conangla}}, \bibinfo {author} {\bibfnamefont {A.~W.}\ \bibnamefont {Schell}},\ and\ \bibinfo {author} {\bibfnamefont {R.}~\bibnamefont {Quidant}},\ }\bibfield  {title} {\bibinfo {title} {Accurate {Mass} {Measurement} of a {Levitated} {Nanomechanical} {Resonator} for {Precision} {Force}-{Sensing}},\ }\href {https://doi.org/10.1021/acs.nanolett.9b00082} {\bibfield  {journal} {\bibinfo  {journal} {Nano Letters}\ }\textbf {\bibinfo {volume} {19}},\ \bibinfo {pages} {6711} (\bibinfo {year} {2019})}\BibitemShut {NoStop}%
\bibitem [{\citenamefont {Slezak}\ \emph {et~al.}(2018)\citenamefont {Slezak}, \citenamefont {Lewandowski}, \citenamefont {Hsu},\ and\ \citenamefont {D’Urso}}]{slezak_cooling_2018}%
  \BibitemOpen
  \bibfield  {author} {\bibinfo {author} {\bibfnamefont {B.~R.}\ \bibnamefont {Slezak}}, \bibinfo {author} {\bibfnamefont {C.~W.}\ \bibnamefont {Lewandowski}}, \bibinfo {author} {\bibfnamefont {J.-F.}\ \bibnamefont {Hsu}},\ and\ \bibinfo {author} {\bibfnamefont {B.}~\bibnamefont {D’Urso}},\ }\bibfield  {title} {\bibinfo {title} {Cooling the motion of a silica microsphere in a magneto-gravitational trap in ultra-high vacuum},\ }\href {https://doi.org/10.1088/1367-2630/aacac1} {\bibfield  {journal} {\bibinfo  {journal} {New Journal of Physics}\ }\textbf {\bibinfo {volume} {20}},\ \bibinfo {pages} {063028} (\bibinfo {year} {2018})}\BibitemShut {NoStop}%
\bibitem [{\citenamefont {Monteiro}\ \emph {et~al.}(2017)\citenamefont {Monteiro}, \citenamefont {Ghosh}, \citenamefont {Fine},\ and\ \citenamefont {Moore}}]{monteiro_optical_2017}%
  \BibitemOpen
  \bibfield  {author} {\bibinfo {author} {\bibfnamefont {F.}~\bibnamefont {Monteiro}}, \bibinfo {author} {\bibfnamefont {S.}~\bibnamefont {Ghosh}}, \bibinfo {author} {\bibfnamefont {A.~G.}\ \bibnamefont {Fine}},\ and\ \bibinfo {author} {\bibfnamefont {D.~C.}\ \bibnamefont {Moore}},\ }\bibfield  {title} {\bibinfo {title} {Optical levitation of 10-ng spheres with nano- g acceleration sensitivity},\ }\href {https://doi.org/10.1103/PhysRevA.96.063841} {\bibfield  {journal} {\bibinfo  {journal} {Physical Review A}\ }\textbf {\bibinfo {volume} {96}},\ \bibinfo {pages} {063841} (\bibinfo {year} {2017})}\BibitemShut {NoStop}%
\bibitem [{\citenamefont {Romero-Isart}\ \emph {et~al.}(2011)\citenamefont {Romero-Isart}, \citenamefont {Pflanzer}, \citenamefont {Blaser}, \citenamefont {Kaltenbaek}, \citenamefont {Kiesel}, \citenamefont {Aspelmeyer},\ and\ \citenamefont {Cirac}}]{romero-isart_large_2011}%
  \BibitemOpen
  \bibfield  {author} {\bibinfo {author} {\bibfnamefont {O.}~\bibnamefont {Romero-Isart}}, \bibinfo {author} {\bibfnamefont {A.~C.}\ \bibnamefont {Pflanzer}}, \bibinfo {author} {\bibfnamefont {F.}~\bibnamefont {Blaser}}, \bibinfo {author} {\bibfnamefont {R.}~\bibnamefont {Kaltenbaek}}, \bibinfo {author} {\bibfnamefont {N.}~\bibnamefont {Kiesel}}, \bibinfo {author} {\bibfnamefont {M.}~\bibnamefont {Aspelmeyer}},\ and\ \bibinfo {author} {\bibfnamefont {J.~I.}\ \bibnamefont {Cirac}},\ }\bibfield  {title} {\bibinfo {title} {Large {Quantum} {Superpositions} and {Interference} of {Massive} {Nanometer}-{Sized} {Objects}},\ }\href {https://doi.org/10.1103/PhysRevLett.107.020405} {\bibfield  {journal} {\bibinfo  {journal} {Physical Review Letters}\ }\textbf {\bibinfo {volume} {107}},\ \bibinfo {pages} {020405} (\bibinfo {year} {2011})}\BibitemShut {NoStop}%
\bibitem [{\citenamefont {Kaltenbaek}\ \emph {et~al.}(2012)\citenamefont {Kaltenbaek}, \citenamefont {Hechenblaikner}, \citenamefont {Kiesel}, \citenamefont {Romero-Isart}, \citenamefont {Schwab}, \citenamefont {Johann},\ and\ \citenamefont {Aspelmeyer}}]{kaltenbaek_macroscopic_2012}%
  \BibitemOpen
  \bibfield  {author} {\bibinfo {author} {\bibfnamefont {R.}~\bibnamefont {Kaltenbaek}}, \bibinfo {author} {\bibfnamefont {G.}~\bibnamefont {Hechenblaikner}}, \bibinfo {author} {\bibfnamefont {N.}~\bibnamefont {Kiesel}}, \bibinfo {author} {\bibfnamefont {O.}~\bibnamefont {Romero-Isart}}, \bibinfo {author} {\bibfnamefont {K.~C.}\ \bibnamefont {Schwab}}, \bibinfo {author} {\bibfnamefont {U.}~\bibnamefont {Johann}},\ and\ \bibinfo {author} {\bibfnamefont {M.}~\bibnamefont {Aspelmeyer}},\ }\bibfield  {title} {\bibinfo {title} {Macroscopic quantum resonators ({MAQRO}): {Testing} quantum and gravitational physics with massive mechanical resonators},\ }\href {https://doi.org/10.1007/s10686-012-9292-3} {\bibfield  {journal} {\bibinfo  {journal} {Experimental Astronomy}\ }\textbf {\bibinfo {volume} {34}},\ \bibinfo {pages} {123} (\bibinfo {year} {2012})}\BibitemShut {NoStop}%
\bibitem [{\citenamefont {Bateman}\ \emph {et~al.}(2014)\citenamefont {Bateman}, \citenamefont {Nimmrichter}, \citenamefont {Hornberger},\ and\ \citenamefont {Ulbricht}}]{bateman_near-field_2014}%
  \BibitemOpen
  \bibfield  {author} {\bibinfo {author} {\bibfnamefont {J.}~\bibnamefont {Bateman}}, \bibinfo {author} {\bibfnamefont {S.}~\bibnamefont {Nimmrichter}}, \bibinfo {author} {\bibfnamefont {K.}~\bibnamefont {Hornberger}},\ and\ \bibinfo {author} {\bibfnamefont {H.}~\bibnamefont {Ulbricht}},\ }\bibfield  {title} {\bibinfo {title} {Near-field interferometry of a free-falling nanoparticle from a point-like source},\ }\href {https://doi.org/10.1038/ncomms5788} {\bibfield  {journal} {\bibinfo  {journal} {Nature Communications}\ }\textbf {\bibinfo {volume} {5}},\ \bibinfo {pages} {4788} (\bibinfo {year} {2014})}\BibitemShut {NoStop}%
\bibitem [{\citenamefont {Wan}\ \emph {et~al.}(2016)\citenamefont {Wan}, \citenamefont {Scala}, \citenamefont {Morley}, \citenamefont {Rahman}, \citenamefont {Ulbricht}, \citenamefont {Bateman}, \citenamefont {Barker}, \citenamefont {Bose},\ and\ \citenamefont {Kim}}]{wan_free_2016}%
  \BibitemOpen
  \bibfield  {author} {\bibinfo {author} {\bibfnamefont {C.}~\bibnamefont {Wan}}, \bibinfo {author} {\bibfnamefont {M.}~\bibnamefont {Scala}}, \bibinfo {author} {\bibfnamefont {G.}~\bibnamefont {Morley}}, \bibinfo {author} {\bibfnamefont {A.~A.}\ \bibnamefont {Rahman}}, \bibinfo {author} {\bibfnamefont {H.}~\bibnamefont {Ulbricht}}, \bibinfo {author} {\bibfnamefont {J.}~\bibnamefont {Bateman}}, \bibinfo {author} {\bibfnamefont {P.}~\bibnamefont {Barker}}, \bibinfo {author} {\bibfnamefont {S.}~\bibnamefont {Bose}},\ and\ \bibinfo {author} {\bibfnamefont {M.}~\bibnamefont {Kim}},\ }\bibfield  {title} {\bibinfo {title} {Free {Nano}-{Object} {Ramsey} {Interferometry} for {Large} {Quantum} {Superpositions}},\ }\href {https://doi.org/10.1103/PhysRevLett.117.143003} {\bibfield  {journal} {\bibinfo  {journal} {Physical Review Letters}\ }\textbf {\bibinfo {volume} {117}},\ \bibinfo {pages} {143003} (\bibinfo {year} {2016})}\BibitemShut {NoStop}%
\bibitem [{\citenamefont {Pino}\ \emph {et~al.}(2018)\citenamefont {Pino}, \citenamefont {Prat-Camps}, \citenamefont {Sinha}, \citenamefont {Venkatesh},\ and\ \citenamefont {Romero-Isart}}]{pino_-chip_2018}%
  \BibitemOpen
  \bibfield  {author} {\bibinfo {author} {\bibfnamefont {H.}~\bibnamefont {Pino}}, \bibinfo {author} {\bibfnamefont {J.}~\bibnamefont {Prat-Camps}}, \bibinfo {author} {\bibfnamefont {K.}~\bibnamefont {Sinha}}, \bibinfo {author} {\bibfnamefont {B.~P.}\ \bibnamefont {Venkatesh}},\ and\ \bibinfo {author} {\bibfnamefont {O.}~\bibnamefont {Romero-Isart}},\ }\bibfield  {title} {\bibinfo {title} {On-chip quantum interference of a superconducting microsphere},\ }\href {https://doi.org/10.1088/2058-9565/aa9d15} {\bibfield  {journal} {\bibinfo  {journal} {Quantum Science and Technology}\ }\textbf {\bibinfo {volume} {3}},\ \bibinfo {pages} {025001} (\bibinfo {year} {2018})}\BibitemShut {NoStop}%
\bibitem [{\citenamefont {Nimmrichter}\ and\ \citenamefont {Hornberger}(2013)}]{nimmrichter_macroscopicity_2013}%
  \BibitemOpen
  \bibfield  {author} {\bibinfo {author} {\bibfnamefont {S.}~\bibnamefont {Nimmrichter}}\ and\ \bibinfo {author} {\bibfnamefont {K.}~\bibnamefont {Hornberger}},\ }\bibfield  {title} {\bibinfo {title} {Macroscopicity of {Mechanical} {Quantum} {Superposition} {States}},\ }\href {https://doi.org/10.1103/PhysRevLett.110.160403} {\bibfield  {journal} {\bibinfo  {journal} {Physical Review Letters}\ }\textbf {\bibinfo {volume} {110}},\ \bibinfo {pages} {160403} (\bibinfo {year} {2013})}\BibitemShut {NoStop}%
\bibitem [{\citenamefont {Bassi}\ \emph {et~al.}(2013)\citenamefont {Bassi}, \citenamefont {Lochan}, \citenamefont {Satin}, \citenamefont {Singh},\ and\ \citenamefont {Ulbricht}}]{bassi_models_2013}%
  \BibitemOpen
  \bibfield  {author} {\bibinfo {author} {\bibfnamefont {A.}~\bibnamefont {Bassi}}, \bibinfo {author} {\bibfnamefont {K.}~\bibnamefont {Lochan}}, \bibinfo {author} {\bibfnamefont {S.}~\bibnamefont {Satin}}, \bibinfo {author} {\bibfnamefont {T.~P.}\ \bibnamefont {Singh}},\ and\ \bibinfo {author} {\bibfnamefont {H.}~\bibnamefont {Ulbricht}},\ }\bibfield  {title} {\bibinfo {title} {Models of wave-function collapse, underlying theories, and experimental tests},\ }\href {https://doi.org/10.1103/RevModPhys.85.471} {\bibfield  {journal} {\bibinfo  {journal} {Reviews of Modern Physics}\ }\textbf {\bibinfo {volume} {85}},\ \bibinfo {pages} {471} (\bibinfo {year} {2013})}\BibitemShut {NoStop}%
\bibitem [{\citenamefont {Bose}\ \emph {et~al.}(2017)\citenamefont {Bose}, \citenamefont {Mazumdar}, \citenamefont {Morley}, \citenamefont {Ulbricht}, \citenamefont {Toroš}, \citenamefont {Paternostro}, \citenamefont {Geraci}, \citenamefont {Barker}, \citenamefont {Kim},\ and\ \citenamefont {Milburn}}]{bose_spin_2017}%
  \BibitemOpen
  \bibfield  {author} {\bibinfo {author} {\bibfnamefont {S.}~\bibnamefont {Bose}}, \bibinfo {author} {\bibfnamefont {A.}~\bibnamefont {Mazumdar}}, \bibinfo {author} {\bibfnamefont {G.~W.}\ \bibnamefont {Morley}}, \bibinfo {author} {\bibfnamefont {H.}~\bibnamefont {Ulbricht}}, \bibinfo {author} {\bibfnamefont {M.}~\bibnamefont {Toroš}}, \bibinfo {author} {\bibfnamefont {M.}~\bibnamefont {Paternostro}}, \bibinfo {author} {\bibfnamefont {A.~A.}\ \bibnamefont {Geraci}}, \bibinfo {author} {\bibfnamefont {P.~F.}\ \bibnamefont {Barker}}, \bibinfo {author} {\bibfnamefont {M.}~\bibnamefont {Kim}},\ and\ \bibinfo {author} {\bibfnamefont {G.}~\bibnamefont {Milburn}},\ }\bibfield  {title} {\bibinfo {title} {Spin {Entanglement} {Witness} for {Quantum} {Gravity}},\ }\href {https://doi.org/10.1103/PhysRevLett.119.240401} {\bibfield  {journal} {\bibinfo  {journal} {Physical Review Letters}\ }\textbf {\bibinfo {volume} {119}},\ \bibinfo {pages} {240401} (\bibinfo {year} {2017})}\BibitemShut {NoStop}%
\bibitem [{\citenamefont {Ashkin}(1970)}]{ashkin_acceleration_1970}%
  \BibitemOpen
  \bibfield  {author} {\bibinfo {author} {\bibfnamefont {A.}~\bibnamefont {Ashkin}},\ }\bibfield  {title} {\bibinfo {title} {Acceleration and {Trapping} of {Particles} by {Radiation} {Pressure}},\ }\href {https://doi.org/10.1103/PhysRevLett.24.156} {\bibfield  {journal} {\bibinfo  {journal} {Physical Review Letters}\ }\textbf {\bibinfo {volume} {24}},\ \bibinfo {pages} {156} (\bibinfo {year} {1970})}\BibitemShut {NoStop}%
\bibitem [{\citenamefont {Chu}\ \emph {et~al.}(1986)\citenamefont {Chu}, \citenamefont {Bjorkholm}, \citenamefont {Ashkin},\ and\ \citenamefont {Cable}}]{chu_experimental_1986}%
  \BibitemOpen
  \bibfield  {author} {\bibinfo {author} {\bibfnamefont {S.}~\bibnamefont {Chu}}, \bibinfo {author} {\bibfnamefont {J.~E.}\ \bibnamefont {Bjorkholm}}, \bibinfo {author} {\bibfnamefont {A.}~\bibnamefont {Ashkin}},\ and\ \bibinfo {author} {\bibfnamefont {A.}~\bibnamefont {Cable}},\ }\bibfield  {title} {\bibinfo {title} {Experimental {Observation} of {Optically} {Trapped} {Atoms}},\ }\href {https://doi.org/10.1103/PhysRevLett.57.314} {\bibfield  {journal} {\bibinfo  {journal} {Physical Review Letters}\ }\textbf {\bibinfo {volume} {57}},\ \bibinfo {pages} {314} (\bibinfo {year} {1986})}\BibitemShut {NoStop}%
\bibitem [{\citenamefont {Chu}(1998)}]{chu_nobel_1998}%
  \BibitemOpen
  \bibfield  {author} {\bibinfo {author} {\bibfnamefont {S.}~\bibnamefont {Chu}},\ }\bibfield  {title} {\bibinfo {title} {Nobel {Lecture}: {The} manipulation of neutral particles},\ }\href {https://doi.org/10.1103/RevModPhys.70.685} {\bibfield  {journal} {\bibinfo  {journal} {Reviews of Modern Physics}\ }\textbf {\bibinfo {volume} {70}},\ \bibinfo {pages} {685} (\bibinfo {year} {1998})}\BibitemShut {NoStop}%
\bibitem [{\citenamefont {Phillips}(1998)}]{phillips_nobel_1998}%
  \BibitemOpen
  \bibfield  {author} {\bibinfo {author} {\bibfnamefont {W.~D.}\ \bibnamefont {Phillips}},\ }\bibfield  {title} {\bibinfo {title} {Nobel {Lecture}: {Laser} cooling and trapping of neutral atoms},\ }\href {https://doi.org/10.1103/RevModPhys.70.721} {\bibfield  {journal} {\bibinfo  {journal} {Reviews of Modern Physics}\ }\textbf {\bibinfo {volume} {70}},\ \bibinfo {pages} {721} (\bibinfo {year} {1998})}\BibitemShut {NoStop}%
\bibitem [{\citenamefont {Hu}\ \emph {et~al.}(2017)\citenamefont {Hu}, \citenamefont {Urvoy}, \citenamefont {Vendeiro}, \citenamefont {Crépel}, \citenamefont {Chen},\ and\ \citenamefont {Vuletić}}]{hu_creation_2017}%
  \BibitemOpen
  \bibfield  {author} {\bibinfo {author} {\bibfnamefont {J.}~\bibnamefont {Hu}}, \bibinfo {author} {\bibfnamefont {A.}~\bibnamefont {Urvoy}}, \bibinfo {author} {\bibfnamefont {Z.}~\bibnamefont {Vendeiro}}, \bibinfo {author} {\bibfnamefont {V.}~\bibnamefont {Crépel}}, \bibinfo {author} {\bibfnamefont {W.}~\bibnamefont {Chen}},\ and\ \bibinfo {author} {\bibfnamefont {V.}~\bibnamefont {Vuletić}},\ }\bibfield  {title} {\bibinfo {title} {Creation of a {Bose}-condensed gas of $^{\textrm{87}}$ {Rb} by laser cooling},\ }\href {https://doi.org/10.1126/science.aan5614} {\bibfield  {journal} {\bibinfo  {journal} {Science}\ }\textbf {\bibinfo {volume} {358}},\ \bibinfo {pages} {1078} (\bibinfo {year} {2017})}\BibitemShut {NoStop}%
\bibitem [{\citenamefont {Moore}\ and\ \citenamefont {Geraci}(2021)}]{moore_searching_2021}%
  \BibitemOpen
  \bibfield  {author} {\bibinfo {author} {\bibfnamefont {D.~C.}\ \bibnamefont {Moore}}\ and\ \bibinfo {author} {\bibfnamefont {A.~A.}\ \bibnamefont {Geraci}},\ }\bibfield  {title} {\bibinfo {title} {Searching for new physics using optically levitated sensors},\ }\href {https://doi.org/10.1088/2058-9565/abcf8a} {\bibfield  {journal} {\bibinfo  {journal} {Quantum Science and Technology}\ }\textbf {\bibinfo {volume} {6}},\ \bibinfo {pages} {014008} (\bibinfo {year} {2021})}\BibitemShut {NoStop}%
\bibitem [{\citenamefont {Ludlow}\ \emph {et~al.}(2015)\citenamefont {Ludlow}, \citenamefont {Boyd}, \citenamefont {Ye}, \citenamefont {Peik},\ and\ \citenamefont {Schmidt}}]{ludlow_optical_2015}%
  \BibitemOpen
  \bibfield  {author} {\bibinfo {author} {\bibfnamefont {A.~D.}\ \bibnamefont {Ludlow}}, \bibinfo {author} {\bibfnamefont {M.~M.}\ \bibnamefont {Boyd}}, \bibinfo {author} {\bibfnamefont {J.}~\bibnamefont {Ye}}, \bibinfo {author} {\bibfnamefont {E.}~\bibnamefont {Peik}},\ and\ \bibinfo {author} {\bibfnamefont {P.}~\bibnamefont {Schmidt}},\ }\bibfield  {title} {\bibinfo {title} {Optical atomic clocks},\ }\href {https://doi.org/10.1103/RevModPhys.87.637} {\bibfield  {journal} {\bibinfo  {journal} {Reviews of Modern Physics}\ }\textbf {\bibinfo {volume} {87}},\ \bibinfo {pages} {637} (\bibinfo {year} {2015})}\BibitemShut {NoStop}%
\bibitem [{\citenamefont {Gieseler}\ \emph {et~al.}(2012)\citenamefont {Gieseler}, \citenamefont {Deutsch}, \citenamefont {Quidant},\ and\ \citenamefont {Novotny}}]{gieseler_subkelvin_2012}%
  \BibitemOpen
  \bibfield  {author} {\bibinfo {author} {\bibfnamefont {J.}~\bibnamefont {Gieseler}}, \bibinfo {author} {\bibfnamefont {B.}~\bibnamefont {Deutsch}}, \bibinfo {author} {\bibfnamefont {R.}~\bibnamefont {Quidant}},\ and\ \bibinfo {author} {\bibfnamefont {L.}~\bibnamefont {Novotny}},\ }\bibfield  {title} {\bibinfo {title} {Subkelvin {Parametric} {Feedback} {Cooling} of a {Laser}-{Trapped} {Nanoparticle}},\ }\href {https://doi.org/10.1103/PhysRevLett.109.103603} {\bibfield  {journal} {\bibinfo  {journal} {Physical Review Letters}\ }\textbf {\bibinfo {volume} {109}},\ \bibinfo {pages} {103603} (\bibinfo {year} {2012})}\BibitemShut {NoStop}%
\bibitem [{\citenamefont {Gao}\ \emph {et~al.}(2024)\citenamefont {Gao}, \citenamefont {van~der Laan}, \citenamefont {Zielińska}, \citenamefont {Militaru}, \citenamefont {Novotny},\ and\ \citenamefont {Frimmer}}]{gao_feedback_2024}%
  \BibitemOpen
  \bibfield  {author} {\bibinfo {author} {\bibfnamefont {J.}~\bibnamefont {Gao}}, \bibinfo {author} {\bibfnamefont {F.}~\bibnamefont {van~der Laan}}, \bibinfo {author} {\bibfnamefont {J.~A.}\ \bibnamefont {Zielińska}}, \bibinfo {author} {\bibfnamefont {A.}~\bibnamefont {Militaru}}, \bibinfo {author} {\bibfnamefont {L.}~\bibnamefont {Novotny}},\ and\ \bibinfo {author} {\bibfnamefont {M.}~\bibnamefont {Frimmer}},\ }\bibfield  {title} {\bibinfo {title} {Feedback cooling a levitated nanoparticle's libration to below 100 phonons},\ }\href {https://doi.org/10.1103/PhysRevResearch.6.033009} {\bibfield  {journal} {\bibinfo  {journal} {Physical Review Research}\ }\textbf {\bibinfo {volume} {6}},\ \bibinfo {pages} {033009} (\bibinfo {year} {2024})}\BibitemShut {NoStop}%
\bibitem [{\citenamefont {Tebbenjohanns}\ \emph {et~al.}(2021)\citenamefont {Tebbenjohanns}, \citenamefont {Mattana}, \citenamefont {Rossi}, \citenamefont {Frimmer},\ and\ \citenamefont {Novotny}}]{tebbenjohanns_quantum_2021}%
  \BibitemOpen
  \bibfield  {author} {\bibinfo {author} {\bibfnamefont {F.}~\bibnamefont {Tebbenjohanns}}, \bibinfo {author} {\bibfnamefont {M.~L.}\ \bibnamefont {Mattana}}, \bibinfo {author} {\bibfnamefont {M.}~\bibnamefont {Rossi}}, \bibinfo {author} {\bibfnamefont {M.}~\bibnamefont {Frimmer}},\ and\ \bibinfo {author} {\bibfnamefont {L.}~\bibnamefont {Novotny}},\ }\bibfield  {title} {\bibinfo {title} {Quantum control of a nanoparticle optically levitated in cryogenic free space},\ }\href {https://doi.org/10.1038/s41586-021-03617-w} {\bibfield  {journal} {\bibinfo  {journal} {Nature}\ }\textbf {\bibinfo {volume} {595}},\ \bibinfo {pages} {378} (\bibinfo {year} {2021})}\BibitemShut {NoStop}%
\bibitem [{\citenamefont {Piotrowski}\ \emph {et~al.}(2023)\citenamefont {Piotrowski}, \citenamefont {Windey}, \citenamefont {Vijayan}, \citenamefont {Gonzalez-Ballestero}, \citenamefont {de~los Ríos~Sommer}, \citenamefont {Meyer}, \citenamefont {Quidant}, \citenamefont {Romero-Isart}, \citenamefont {Reimann},\ and\ \citenamefont {Novotny}}]{piotrowski_simultaneous_2023}%
  \BibitemOpen
  \bibfield  {author} {\bibinfo {author} {\bibfnamefont {J.}~\bibnamefont {Piotrowski}}, \bibinfo {author} {\bibfnamefont {D.}~\bibnamefont {Windey}}, \bibinfo {author} {\bibfnamefont {J.}~\bibnamefont {Vijayan}}, \bibinfo {author} {\bibfnamefont {C.}~\bibnamefont {Gonzalez-Ballestero}}, \bibinfo {author} {\bibfnamefont {A.}~\bibnamefont {de~los Ríos~Sommer}}, \bibinfo {author} {\bibfnamefont {N.}~\bibnamefont {Meyer}}, \bibinfo {author} {\bibfnamefont {R.}~\bibnamefont {Quidant}}, \bibinfo {author} {\bibfnamefont {O.}~\bibnamefont {Romero-Isart}}, \bibinfo {author} {\bibfnamefont {R.}~\bibnamefont {Reimann}},\ and\ \bibinfo {author} {\bibfnamefont {L.}~\bibnamefont {Novotny}},\ }\bibfield  {title} {\bibinfo {title} {Simultaneous ground-state cooling of two mechanical modes of a levitated nanoparticle},\ }\href {https://doi.org/10.1038/s41567-023-01956-1} {\bibfield  {journal} {\bibinfo  {journal} {Nature Physics}\ }\textbf {\bibinfo {volume} {19}},\ \bibinfo {pages} {1009} (\bibinfo {year}
  {2023})}\BibitemShut {NoStop}%
\bibitem [{\citenamefont {Rahman}\ \emph {et~al.}(2016)\citenamefont {Rahman}, \citenamefont {Frangeskou}, \citenamefont {Kim}, \citenamefont {Bose}, \citenamefont {Morley},\ and\ \citenamefont {Barker}}]{rahman_burning_2016}%
  \BibitemOpen
  \bibfield  {author} {\bibinfo {author} {\bibfnamefont {A.~T. M.~A.}\ \bibnamefont {Rahman}}, \bibinfo {author} {\bibfnamefont {A.~C.}\ \bibnamefont {Frangeskou}}, \bibinfo {author} {\bibfnamefont {M.~S.}\ \bibnamefont {Kim}}, \bibinfo {author} {\bibfnamefont {S.}~\bibnamefont {Bose}}, \bibinfo {author} {\bibfnamefont {G.~W.}\ \bibnamefont {Morley}},\ and\ \bibinfo {author} {\bibfnamefont {P.~F.}\ \bibnamefont {Barker}},\ }\bibfield  {title} {\bibinfo {title} {Burning and graphitization of optically levitated nanodiamonds in vacuum},\ }\href {https://doi.org/10.1038/srep21633} {\bibfield  {journal} {\bibinfo  {journal} {Scientific Reports}\ }\textbf {\bibinfo {volume} {6}},\ \bibinfo {pages} {21633} (\bibinfo {year} {2016})}\BibitemShut {NoStop}%
\bibitem [{\citenamefont {Paul}(1990)}]{paul_electromagnetic_1990}%
  \BibitemOpen
  \bibfield  {author} {\bibinfo {author} {\bibfnamefont {W.}~\bibnamefont {Paul}},\ }\bibfield  {title} {\bibinfo {title} {Electromagnetic traps for charged and neutral particles},\ }\href {https://doi.org/10.1103/RevModPhys.62.531} {\bibfield  {journal} {\bibinfo  {journal} {Reviews of Modern Physics}\ }\textbf {\bibinfo {volume} {62}},\ \bibinfo {pages} {531} (\bibinfo {year} {1990})}\BibitemShut {NoStop}%
\bibitem [{\citenamefont {Ostermayr}\ \emph {et~al.}(2018)\citenamefont {Ostermayr}, \citenamefont {Gebhard}, \citenamefont {Haffa}, \citenamefont {Kiefer}, \citenamefont {Kreuzer}, \citenamefont {Allinger}, \citenamefont {Bömer}, \citenamefont {Braenzel}, \citenamefont {Schnürer}, \citenamefont {Cermak}, \citenamefont {Schreiber},\ and\ \citenamefont {Hilz}}]{ostermayr_transportable_2018}%
  \BibitemOpen
  \bibfield  {author} {\bibinfo {author} {\bibfnamefont {T.~M.}\ \bibnamefont {Ostermayr}}, \bibinfo {author} {\bibfnamefont {J.}~\bibnamefont {Gebhard}}, \bibinfo {author} {\bibfnamefont {D.}~\bibnamefont {Haffa}}, \bibinfo {author} {\bibfnamefont {D.}~\bibnamefont {Kiefer}}, \bibinfo {author} {\bibfnamefont {C.}~\bibnamefont {Kreuzer}}, \bibinfo {author} {\bibfnamefont {K.}~\bibnamefont {Allinger}}, \bibinfo {author} {\bibfnamefont {C.}~\bibnamefont {Bömer}}, \bibinfo {author} {\bibfnamefont {J.}~\bibnamefont {Braenzel}}, \bibinfo {author} {\bibfnamefont {M.}~\bibnamefont {Schnürer}}, \bibinfo {author} {\bibfnamefont {I.}~\bibnamefont {Cermak}}, \bibinfo {author} {\bibfnamefont {J.}~\bibnamefont {Schreiber}},\ and\ \bibinfo {author} {\bibfnamefont {P.}~\bibnamefont {Hilz}},\ }\bibfield  {title} {\bibinfo {title} {A transportable {Paul}-trap for levitation and accurate positioning of micron-scale particles in vacuum for laser-plasma experiments},\ }\href {https://doi.org/10.1063/1.4995955} {\bibfield
  {journal} {\bibinfo  {journal} {Review of Scientific Instruments}\ }\textbf {\bibinfo {volume} {89}},\ \bibinfo {pages} {013302} (\bibinfo {year} {2018})}\BibitemShut {NoStop}%
\bibitem [{\citenamefont {Dania}\ \emph {et~al.}(2021)\citenamefont {Dania}, \citenamefont {Bykov}, \citenamefont {Knoll}, \citenamefont {Mestres},\ and\ \citenamefont {Northup}}]{dania_optical_2021}%
  \BibitemOpen
  \bibfield  {author} {\bibinfo {author} {\bibfnamefont {L.}~\bibnamefont {Dania}}, \bibinfo {author} {\bibfnamefont {D.~S.}\ \bibnamefont {Bykov}}, \bibinfo {author} {\bibfnamefont {M.}~\bibnamefont {Knoll}}, \bibinfo {author} {\bibfnamefont {P.}~\bibnamefont {Mestres}},\ and\ \bibinfo {author} {\bibfnamefont {T.~E.}\ \bibnamefont {Northup}},\ }\bibfield  {title} {\bibinfo {title} {Optical and electrical feedback cooling of a silica nanoparticle levitated in a {Paul} trap},\ }\href {https://doi.org/10.1103/PhysRevResearch.3.013018} {\bibfield  {journal} {\bibinfo  {journal} {Physical Review Research}\ }\textbf {\bibinfo {volume} {3}},\ \bibinfo {pages} {013018} (\bibinfo {year} {2021})}\BibitemShut {NoStop}%
\bibitem [{\citenamefont {Jain}\ \emph {et~al.}(2024)\citenamefont {Jain}, \citenamefont {Sägesser}, \citenamefont {Hrmo}, \citenamefont {Torkzaban}, \citenamefont {Stadler}, \citenamefont {Oswald}, \citenamefont {Axline}, \citenamefont {Bautista-Salvador}, \citenamefont {Ospelkaus}, \citenamefont {Kienzler},\ and\ \citenamefont {Home}}]{jain_penning_2024}%
  \BibitemOpen
  \bibfield  {author} {\bibinfo {author} {\bibfnamefont {S.}~\bibnamefont {Jain}}, \bibinfo {author} {\bibfnamefont {T.}~\bibnamefont {Sägesser}}, \bibinfo {author} {\bibfnamefont {P.}~\bibnamefont {Hrmo}}, \bibinfo {author} {\bibfnamefont {C.}~\bibnamefont {Torkzaban}}, \bibinfo {author} {\bibfnamefont {M.}~\bibnamefont {Stadler}}, \bibinfo {author} {\bibfnamefont {R.}~\bibnamefont {Oswald}}, \bibinfo {author} {\bibfnamefont {C.}~\bibnamefont {Axline}}, \bibinfo {author} {\bibfnamefont {A.}~\bibnamefont {Bautista-Salvador}}, \bibinfo {author} {\bibfnamefont {C.}~\bibnamefont {Ospelkaus}}, \bibinfo {author} {\bibfnamefont {D.}~\bibnamefont {Kienzler}},\ and\ \bibinfo {author} {\bibfnamefont {J.}~\bibnamefont {Home}},\ }\bibfield  {title} {\bibinfo {title} {Penning micro-trap for quantum computing},\ }\href {https://doi.org/10.1038/s41586-024-07111-x} {\bibfield  {journal} {\bibinfo  {journal} {Nature}\ }\textbf {\bibinfo {volume} {627}},\ \bibinfo {pages} {510} (\bibinfo {year} {2024})}\BibitemShut {NoStop}%
\bibitem [{\citenamefont {Delord}\ \emph {et~al.}(2020)\citenamefont {Delord}, \citenamefont {Huillery}, \citenamefont {Nicolas},\ and\ \citenamefont {Hétet}}]{delord_spin-cooling_2020}%
  \BibitemOpen
  \bibfield  {author} {\bibinfo {author} {\bibfnamefont {T.}~\bibnamefont {Delord}}, \bibinfo {author} {\bibfnamefont {P.}~\bibnamefont {Huillery}}, \bibinfo {author} {\bibfnamefont {L.}~\bibnamefont {Nicolas}},\ and\ \bibinfo {author} {\bibfnamefont {G.}~\bibnamefont {Hétet}},\ }\bibfield  {title} {\bibinfo {title} {Spin-cooling of the motion of a trapped diamond},\ }\href {https://doi.org/10.1038/s41586-020-2133-z} {\bibfield  {journal} {\bibinfo  {journal} {Nature}\ }\textbf {\bibinfo {volume} {580}},\ \bibinfo {pages} {56} (\bibinfo {year} {2020})}\BibitemShut {NoStop}%
\bibitem [{\citenamefont {Alda}\ \emph {et~al.}(2016)\citenamefont {Alda}, \citenamefont {Berthelot}, \citenamefont {Rica},\ and\ \citenamefont {Quidant}}]{alda_trapping_2016}%
  \BibitemOpen
  \bibfield  {author} {\bibinfo {author} {\bibfnamefont {I.}~\bibnamefont {Alda}}, \bibinfo {author} {\bibfnamefont {J.}~\bibnamefont {Berthelot}}, \bibinfo {author} {\bibfnamefont {R.~A.}\ \bibnamefont {Rica}},\ and\ \bibinfo {author} {\bibfnamefont {R.}~\bibnamefont {Quidant}},\ }\bibfield  {title} {\bibinfo {title} {Trapping and manipulation of individual nanoparticles in a planar {Paul} trap},\ }\href {https://doi.org/10.1063/1.4965859} {\bibfield  {journal} {\bibinfo  {journal} {Applied Physics Letters}\ }\textbf {\bibinfo {volume} {109}},\ \bibinfo {pages} {163105} (\bibinfo {year} {2016})}\BibitemShut {NoStop}%
\bibitem [{\citenamefont {Matthiesen}\ \emph {et~al.}(2021)\citenamefont {Matthiesen}, \citenamefont {Yu}, \citenamefont {Guo}, \citenamefont {Alonso},\ and\ \citenamefont {Häffner}}]{matthiesen_trapping_2021}%
  \BibitemOpen
  \bibfield  {author} {\bibinfo {author} {\bibfnamefont {C.}~\bibnamefont {Matthiesen}}, \bibinfo {author} {\bibfnamefont {Q.}~\bibnamefont {Yu}}, \bibinfo {author} {\bibfnamefont {J.}~\bibnamefont {Guo}}, \bibinfo {author} {\bibfnamefont {A.~M.}\ \bibnamefont {Alonso}},\ and\ \bibinfo {author} {\bibfnamefont {H.}~\bibnamefont {Häffner}},\ }\bibfield  {title} {\bibinfo {title} {Trapping {Electrons} in a {Room}-{Temperature} {Microwave} {Paul} {Trap}},\ }\href {https://doi.org/10.1103/PhysRevX.11.011019} {\bibfield  {journal} {\bibinfo  {journal} {Physical Review X}\ }\textbf {\bibinfo {volume} {11}},\ \bibinfo {pages} {011019} (\bibinfo {year} {2021})}\BibitemShut {NoStop}%
\bibitem [{\citenamefont {Hofer}\ \emph {et~al.}(2023)\citenamefont {Hofer}, \citenamefont {Gross}, \citenamefont {Higgins}, \citenamefont {Huebl}, \citenamefont {Kieler}, \citenamefont {Kleiner}, \citenamefont {Koelle}, \citenamefont {Schmidt}, \citenamefont {Slater}, \citenamefont {Trupke}, \citenamefont {Uhl}, \citenamefont {Weimann}, \citenamefont {Wieczorek},\ and\ \citenamefont {Aspelmeyer}}]{hofer_high-_2023}%
  \BibitemOpen
  \bibfield  {author} {\bibinfo {author} {\bibfnamefont {J.}~\bibnamefont {Hofer}}, \bibinfo {author} {\bibfnamefont {R.}~\bibnamefont {Gross}}, \bibinfo {author} {\bibfnamefont {G.}~\bibnamefont {Higgins}}, \bibinfo {author} {\bibfnamefont {H.}~\bibnamefont {Huebl}}, \bibinfo {author} {\bibfnamefont {O.}~\bibnamefont {Kieler}}, \bibinfo {author} {\bibfnamefont {R.}~\bibnamefont {Kleiner}}, \bibinfo {author} {\bibfnamefont {D.}~\bibnamefont {Koelle}}, \bibinfo {author} {\bibfnamefont {P.}~\bibnamefont {Schmidt}}, \bibinfo {author} {\bibfnamefont {J.}~\bibnamefont {Slater}}, \bibinfo {author} {\bibfnamefont {M.}~\bibnamefont {Trupke}}, \bibinfo {author} {\bibfnamefont {K.}~\bibnamefont {Uhl}}, \bibinfo {author} {\bibfnamefont {T.}~\bibnamefont {Weimann}}, \bibinfo {author} {\bibfnamefont {W.}~\bibnamefont {Wieczorek}},\ and\ \bibinfo {author} {\bibfnamefont {M.}~\bibnamefont {Aspelmeyer}},\ }\bibfield  {title} {\bibinfo {title} {High- {Q} {Magnetic} {Levitation} and {Control} of {Superconducting}
  {Microspheres} at {Millikelvin} {Temperatures}},\ }\href {https://doi.org/10.1103/PhysRevLett.131.043603} {\bibfield  {journal} {\bibinfo  {journal} {Physical Review Letters}\ }\textbf {\bibinfo {volume} {131}},\ \bibinfo {pages} {043603} (\bibinfo {year} {2023})}\BibitemShut {NoStop}%
\bibitem [{\citenamefont {Gutierrez~Latorre}\ \emph {et~al.}(2023)\citenamefont {Gutierrez~Latorre}, \citenamefont {Higgins}, \citenamefont {Paradkar}, \citenamefont {Bauch},\ and\ \citenamefont {Wieczorek}}]{gutierrez_latorre_superconducting_2023}%
  \BibitemOpen
  \bibfield  {author} {\bibinfo {author} {\bibfnamefont {M.}~\bibnamefont {Gutierrez~Latorre}}, \bibinfo {author} {\bibfnamefont {G.}~\bibnamefont {Higgins}}, \bibinfo {author} {\bibfnamefont {A.}~\bibnamefont {Paradkar}}, \bibinfo {author} {\bibfnamefont {T.}~\bibnamefont {Bauch}},\ and\ \bibinfo {author} {\bibfnamefont {W.}~\bibnamefont {Wieczorek}},\ }\bibfield  {title} {\bibinfo {title} {Superconducting {Microsphere} {Magnetically} {Levitated} in an {Anharmonic} {Potential} with {Integrated} {Magnetic} {Readout}},\ }\href {https://doi.org/10.1103/PhysRevApplied.19.054047} {\bibfield  {journal} {\bibinfo  {journal} {Physical Review Applied}\ }\textbf {\bibinfo {volume} {19}},\ \bibinfo {pages} {054047} (\bibinfo {year} {2023})}\BibitemShut {NoStop}%
\bibitem [{\citenamefont {Fuchs}\ \emph {et~al.}(2024)\citenamefont {Fuchs}, \citenamefont {Uitenbroek}, \citenamefont {Plugge}, \citenamefont {van Halteren}, \citenamefont {van Soest}, \citenamefont {Vinante}, \citenamefont {Ulbricht},\ and\ \citenamefont {Oosterkamp}}]{fuchs_measuring_2024}%
  \BibitemOpen
  \bibfield  {author} {\bibinfo {author} {\bibfnamefont {T.~M.}\ \bibnamefont {Fuchs}}, \bibinfo {author} {\bibfnamefont {D.~G.}\ \bibnamefont {Uitenbroek}}, \bibinfo {author} {\bibfnamefont {J.}~\bibnamefont {Plugge}}, \bibinfo {author} {\bibfnamefont {N.}~\bibnamefont {van Halteren}}, \bibinfo {author} {\bibfnamefont {J.-P.}\ \bibnamefont {van Soest}}, \bibinfo {author} {\bibfnamefont {A.}~\bibnamefont {Vinante}}, \bibinfo {author} {\bibfnamefont {H.}~\bibnamefont {Ulbricht}},\ and\ \bibinfo {author} {\bibfnamefont {T.~H.}\ \bibnamefont {Oosterkamp}},\ }\bibfield  {title} {\bibinfo {title} {Measuring gravity with milligram levitated masses},\ }\href {https://doi.org/10.1126/sciadv.adk2949} {\bibfield  {journal} {\bibinfo  {journal} {Science Advances}\ }\textbf {\bibinfo {volume} {10}},\ \bibinfo {pages} {eadk2949} (\bibinfo {year} {2024})}\BibitemShut {NoStop}%
\bibitem [{\citenamefont {Timberlake}\ \emph {et~al.}(2019)\citenamefont {Timberlake}, \citenamefont {Gasbarri}, \citenamefont {Vinante}, \citenamefont {Setter},\ and\ \citenamefont {Ulbricht}}]{timberlake_acceleration_2019}%
  \BibitemOpen
  \bibfield  {author} {\bibinfo {author} {\bibfnamefont {C.}~\bibnamefont {Timberlake}}, \bibinfo {author} {\bibfnamefont {G.}~\bibnamefont {Gasbarri}}, \bibinfo {author} {\bibfnamefont {A.}~\bibnamefont {Vinante}}, \bibinfo {author} {\bibfnamefont {A.}~\bibnamefont {Setter}},\ and\ \bibinfo {author} {\bibfnamefont {H.}~\bibnamefont {Ulbricht}},\ }\bibfield  {title} {\bibinfo {title} {Acceleration sensing with magnetically levitated oscillators above a superconductor},\ }\href {https://doi.org/10.1063/1.5129145} {\bibfield  {journal} {\bibinfo  {journal} {Applied Physics Letters}\ }\textbf {\bibinfo {volume} {115}},\ \bibinfo {pages} {224101} (\bibinfo {year} {2019})}\BibitemShut {NoStop}%
\bibitem [{\citenamefont {Timberlake}\ \emph {et~al.}(2023)\citenamefont {Timberlake}, \citenamefont {Simcox},\ and\ \citenamefont {Ulbricht}}]{timberlake_linear_2023}%
  \BibitemOpen
  \bibfield  {author} {\bibinfo {author} {\bibfnamefont {C.}~\bibnamefont {Timberlake}}, \bibinfo {author} {\bibfnamefont {E.}~\bibnamefont {Simcox}},\ and\ \bibinfo {author} {\bibfnamefont {H.}~\bibnamefont {Ulbricht}},\ }\href {https://doi.org/10.48550/ARXIV.2310.03880} {\bibinfo {title} {Linear cooling of a levitated micromagnetic cylinder by vibration}} (\bibinfo {year} {2023}),\ \bibinfo {note} {version Number: 2}\BibitemShut {NoStop}%
\bibitem [{\citenamefont {Vinante}\ \emph {et~al.}(2020)\citenamefont {Vinante}, \citenamefont {Falferi}, \citenamefont {Gasbarri}, \citenamefont {Setter}, \citenamefont {Timberlake},\ and\ \citenamefont {Ulbricht}}]{vinante_ultralow_2020}%
  \BibitemOpen
  \bibfield  {author} {\bibinfo {author} {\bibfnamefont {A.}~\bibnamefont {Vinante}}, \bibinfo {author} {\bibfnamefont {P.}~\bibnamefont {Falferi}}, \bibinfo {author} {\bibfnamefont {G.}~\bibnamefont {Gasbarri}}, \bibinfo {author} {\bibfnamefont {A.}~\bibnamefont {Setter}}, \bibinfo {author} {\bibfnamefont {C.}~\bibnamefont {Timberlake}},\ and\ \bibinfo {author} {\bibfnamefont {H.}~\bibnamefont {Ulbricht}},\ }\bibfield  {title} {\bibinfo {title} {Ultralow {Mechanical} {Damping} with {Meissner}-{Levitated} {Ferromagnetic} {Microparticles}},\ }\href {https://doi.org/10.1103/PhysRevApplied.13.064027} {\bibfield  {journal} {\bibinfo  {journal} {Physical Review Applied}\ }\textbf {\bibinfo {volume} {13}},\ \bibinfo {pages} {064027} (\bibinfo {year} {2020})}\BibitemShut {NoStop}%
\bibitem [{\citenamefont {Gieseler}\ \emph {et~al.}(2020)\citenamefont {Gieseler}, \citenamefont {Kabcenell}, \citenamefont {Rosenfeld}, \citenamefont {Schaefer}, \citenamefont {Safira}, \citenamefont {Schuetz}, \citenamefont {Gonzalez-Ballestero}, \citenamefont {Rusconi}, \citenamefont {Romero-Isart},\ and\ \citenamefont {Lukin}}]{gieseler_single-spin_2020}%
  \BibitemOpen
  \bibfield  {author} {\bibinfo {author} {\bibfnamefont {J.}~\bibnamefont {Gieseler}}, \bibinfo {author} {\bibfnamefont {A.}~\bibnamefont {Kabcenell}}, \bibinfo {author} {\bibfnamefont {E.}~\bibnamefont {Rosenfeld}}, \bibinfo {author} {\bibfnamefont {J.}~\bibnamefont {Schaefer}}, \bibinfo {author} {\bibfnamefont {A.}~\bibnamefont {Safira}}, \bibinfo {author} {\bibfnamefont {M.}~\bibnamefont {Schuetz}}, \bibinfo {author} {\bibfnamefont {C.}~\bibnamefont {Gonzalez-Ballestero}}, \bibinfo {author} {\bibfnamefont {C.}~\bibnamefont {Rusconi}}, \bibinfo {author} {\bibfnamefont {O.}~\bibnamefont {Romero-Isart}},\ and\ \bibinfo {author} {\bibfnamefont {M.}~\bibnamefont {Lukin}},\ }\bibfield  {title} {\bibinfo {title} {Single-{Spin} {Magnetomechanics} with {Levitated} {Micromagnets}},\ }\href {https://doi.org/10.1103/PhysRevLett.124.163604} {\bibfield  {journal} {\bibinfo  {journal} {Physical Review Letters}\ }\textbf {\bibinfo {volume} {124}},\ \bibinfo {pages} {163604} (\bibinfo {year} {2020})}\BibitemShut {NoStop}%
\bibitem [{\citenamefont {Hsu}\ \emph {et~al.}(2016)\citenamefont {Hsu}, \citenamefont {Ji}, \citenamefont {Lewandowski},\ and\ \citenamefont {D’Urso}}]{hsu_cooling_2016}%
  \BibitemOpen
  \bibfield  {author} {\bibinfo {author} {\bibfnamefont {J.-F.}\ \bibnamefont {Hsu}}, \bibinfo {author} {\bibfnamefont {P.}~\bibnamefont {Ji}}, \bibinfo {author} {\bibfnamefont {C.~W.}\ \bibnamefont {Lewandowski}},\ and\ \bibinfo {author} {\bibfnamefont {B.}~\bibnamefont {D’Urso}},\ }\bibfield  {title} {\bibinfo {title} {Cooling the {Motion} of {Diamond} {Nanocrystals} in a {Magneto}-{Gravitational} {Trap} in {High} {Vacuum}},\ }\href {https://doi.org/10.1038/srep30125} {\bibfield  {journal} {\bibinfo  {journal} {Scientific Reports}\ }\textbf {\bibinfo {volume} {6}},\ \bibinfo {pages} {30125} (\bibinfo {year} {2016})}\BibitemShut {NoStop}%
\bibitem [{\citenamefont {Barry}\ \emph {et~al.}(2014)\citenamefont {Barry}, \citenamefont {McCarron}, \citenamefont {Norrgard}, \citenamefont {Steinecker},\ and\ \citenamefont {DeMille}}]{barry_magneto-optical_2014}%
  \BibitemOpen
  \bibfield  {author} {\bibinfo {author} {\bibfnamefont {J.~F.}\ \bibnamefont {Barry}}, \bibinfo {author} {\bibfnamefont {D.~J.}\ \bibnamefont {McCarron}}, \bibinfo {author} {\bibfnamefont {E.~B.}\ \bibnamefont {Norrgard}}, \bibinfo {author} {\bibfnamefont {M.~H.}\ \bibnamefont {Steinecker}},\ and\ \bibinfo {author} {\bibfnamefont {D.}~\bibnamefont {DeMille}},\ }\bibfield  {title} {\bibinfo {title} {Magneto-optical trapping of a diatomic molecule},\ }\href {https://doi.org/10.1038/nature13634} {\bibfield  {journal} {\bibinfo  {journal} {Nature}\ }\textbf {\bibinfo {volume} {512}},\ \bibinfo {pages} {286} (\bibinfo {year} {2014})}\BibitemShut {NoStop}%
\bibitem [{\citenamefont {Hermansen}\ \emph {et~al.}(2023)\citenamefont {Hermansen}, \citenamefont {Durhuus}, \citenamefont {Frandsen}, \citenamefont {Beleggia}, \citenamefont {Bahl},\ and\ \citenamefont {Bjørk}}]{hermansen_magnetic_2023}%
  \BibitemOpen
  \bibfield  {author} {\bibinfo {author} {\bibfnamefont {J.~M.}\ \bibnamefont {Hermansen}}, \bibinfo {author} {\bibfnamefont {F.~L.}\ \bibnamefont {Durhuus}}, \bibinfo {author} {\bibfnamefont {C.}~\bibnamefont {Frandsen}}, \bibinfo {author} {\bibfnamefont {M.}~\bibnamefont {Beleggia}}, \bibinfo {author} {\bibfnamefont {C.~R.}\ \bibnamefont {Bahl}},\ and\ \bibinfo {author} {\bibfnamefont {R.}~\bibnamefont {Bjørk}},\ }\bibfield  {title} {\bibinfo {title} {Magnetic levitation by rotation},\ }\href {https://doi.org/10.1103/PhysRevApplied.20.044036} {\bibfield  {journal} {\bibinfo  {journal} {Physical Review Applied}\ }\textbf {\bibinfo {volume} {20}},\ \bibinfo {pages} {044036} (\bibinfo {year} {2023})}\BibitemShut {NoStop}%
\bibitem [{\citenamefont {Sakuma}(2023)}]{sakuma_three-dimensional_2023}%
  \BibitemOpen
  \bibfield  {author} {\bibinfo {author} {\bibfnamefont {H.}~\bibnamefont {Sakuma}},\ }\bibfield  {title} {\bibinfo {title} {Three-dimensional motion control of an untethered magnetic object using three rotating permanent magnets},\ }\href {https://doi.org/10.1038/s41598-023-45419-2} {\bibfield  {journal} {\bibinfo  {journal} {Scientific Reports}\ }\textbf {\bibinfo {volume} {13}},\ \bibinfo {pages} {18052} (\bibinfo {year} {2023})}\BibitemShut {NoStop}%
\bibitem [{\citenamefont {Simon}\ \emph {et~al.}(1997)\citenamefont {Simon}, \citenamefont {Heflinger},\ and\ \citenamefont {Ridgway}}]{simon_spin_1997}%
  \BibitemOpen
  \bibfield  {author} {\bibinfo {author} {\bibfnamefont {M.~D.}\ \bibnamefont {Simon}}, \bibinfo {author} {\bibfnamefont {L.~O.}\ \bibnamefont {Heflinger}},\ and\ \bibinfo {author} {\bibfnamefont {S.~L.}\ \bibnamefont {Ridgway}},\ }\bibfield  {title} {\bibinfo {title} {Spin stabilized magnetic levitation},\ }\href {https://doi.org/10.1119/1.18488} {\bibfield  {journal} {\bibinfo  {journal} {American Journal of Physics}\ }\textbf {\bibinfo {volume} {65}},\ \bibinfo {pages} {286} (\bibinfo {year} {1997})}\BibitemShut {NoStop}%
\bibitem [{\citenamefont {Sackett}\ \emph {et~al.}(1993)\citenamefont {Sackett}, \citenamefont {Cornell}, \citenamefont {Monroe},\ and\ \citenamefont {Wieman}}]{sackett_magnetic_1993}%
  \BibitemOpen
  \bibfield  {author} {\bibinfo {author} {\bibfnamefont {C.}~\bibnamefont {Sackett}}, \bibinfo {author} {\bibfnamefont {E.}~\bibnamefont {Cornell}}, \bibinfo {author} {\bibfnamefont {C.}~\bibnamefont {Monroe}},\ and\ \bibinfo {author} {\bibfnamefont {C.}~\bibnamefont {Wieman}},\ }\bibfield  {title} {\bibinfo {title} {A magnetic suspension system for atoms and bar magnets},\ }\href {https://doi.org/10.1119/1.17261} {\bibfield  {journal} {\bibinfo  {journal} {American Journal of Physics}\ }\textbf {\bibinfo {volume} {61}},\ \bibinfo {pages} {304} (\bibinfo {year} {1993})}\BibitemShut {NoStop}%
\bibitem [{\citenamefont {Cornell}\ \emph {et~al.}(1991)\citenamefont {Cornell}, \citenamefont {Monroe},\ and\ \citenamefont {Wieman}}]{cornell_multiply_1991}%
  \BibitemOpen
  \bibfield  {author} {\bibinfo {author} {\bibfnamefont {E.~A.}\ \bibnamefont {Cornell}}, \bibinfo {author} {\bibfnamefont {C.}~\bibnamefont {Monroe}},\ and\ \bibinfo {author} {\bibfnamefont {C.~E.}\ \bibnamefont {Wieman}},\ }\bibfield  {title} {\bibinfo {title} {Multiply loaded, ac magnetic trap for neutral atoms},\ }\href {https://doi.org/10.1103/PhysRevLett.67.2439} {\bibfield  {journal} {\bibinfo  {journal} {Physical Review Letters}\ }\textbf {\bibinfo {volume} {67}},\ \bibinfo {pages} {2439} (\bibinfo {year} {1991})}\BibitemShut {NoStop}%
\bibitem [{\citenamefont {Perdriat}\ \emph {et~al.}(2023)\citenamefont {Perdriat}, \citenamefont {Pellet-Mary}, \citenamefont {Copie},\ and\ \citenamefont {Hétet}}]{perdriat_planar_2023}%
  \BibitemOpen
  \bibfield  {author} {\bibinfo {author} {\bibfnamefont {M.}~\bibnamefont {Perdriat}}, \bibinfo {author} {\bibfnamefont {C.}~\bibnamefont {Pellet-Mary}}, \bibinfo {author} {\bibfnamefont {T.}~\bibnamefont {Copie}},\ and\ \bibinfo {author} {\bibfnamefont {G.}~\bibnamefont {Hétet}},\ }\bibfield  {title} {\bibinfo {title} {Planar magnetic {Paul} traps for ferromagnetic particles},\ }\href {https://doi.org/10.1103/PhysRevResearch.5.L032045} {\bibfield  {journal} {\bibinfo  {journal} {Physical Review Research}\ }\textbf {\bibinfo {volume} {5}},\ \bibinfo {pages} {L032045} (\bibinfo {year} {2023})}\BibitemShut {NoStop}%
\bibitem [{\citenamefont {Dehmelt}(1968)}]{dehmelt_radiofrequency_1968}%
  \BibitemOpen
  \bibfield  {author} {\bibinfo {author} {\bibfnamefont {H.}~\bibnamefont {Dehmelt}},\ }\bibfield  {title} {\bibinfo {title} {Radiofrequency {Spectroscopy} of {Stored} {Ions} {I}: {Storage}},\ }in\ \href {https://doi.org/10.1016/S0065-2199(08)60170-0} {\emph {\bibinfo {booktitle} {Advances in {Atomic} and {Molecular} {Physics}}}},\ Vol.~\bibinfo {volume} {3}\ (\bibinfo  {publisher} {Elsevier},\ \bibinfo {year} {1968})\ pp.\ \bibinfo {pages} {53--72}\BibitemShut {NoStop}%
\bibitem [{\citenamefont {Ranjit}\ \emph {et~al.}(2016)\citenamefont {Ranjit}, \citenamefont {Cunningham}, \citenamefont {Casey},\ and\ \citenamefont {Geraci}}]{ranjit_zeptonewton_2016}%
  \BibitemOpen
  \bibfield  {author} {\bibinfo {author} {\bibfnamefont {G.}~\bibnamefont {Ranjit}}, \bibinfo {author} {\bibfnamefont {M.}~\bibnamefont {Cunningham}}, \bibinfo {author} {\bibfnamefont {K.}~\bibnamefont {Casey}},\ and\ \bibinfo {author} {\bibfnamefont {A.~A.}\ \bibnamefont {Geraci}},\ }\bibfield  {title} {\bibinfo {title} {Zeptonewton force sensing with nanospheres in an optical lattice},\ }\href {https://doi.org/10.1103/PhysRevA.93.053801} {\bibfield  {journal} {\bibinfo  {journal} {Physical Review A}\ }\textbf {\bibinfo {volume} {93}},\ \bibinfo {pages} {053801} (\bibinfo {year} {2016})}\BibitemShut {NoStop}%
\bibitem [{\citenamefont {Schmöle}\ \emph {et~al.}(2016)\citenamefont {Schmöle}, \citenamefont {Dragosits}, \citenamefont {Hepach},\ and\ \citenamefont {Aspelmeyer}}]{schmole_micromechanical_2016}%
  \BibitemOpen
  \bibfield  {author} {\bibinfo {author} {\bibfnamefont {J.}~\bibnamefont {Schmöle}}, \bibinfo {author} {\bibfnamefont {M.}~\bibnamefont {Dragosits}}, \bibinfo {author} {\bibfnamefont {H.}~\bibnamefont {Hepach}},\ and\ \bibinfo {author} {\bibfnamefont {M.}~\bibnamefont {Aspelmeyer}},\ }\bibfield  {title} {\bibinfo {title} {A micromechanical proof-of-principle experiment for measuring the gravitational force of milligram masses},\ }\href {https://doi.org/10.1088/0264-9381/33/12/125031} {\bibfield  {journal} {\bibinfo  {journal} {Classical and Quantum Gravity}\ }\textbf {\bibinfo {volume} {33}},\ \bibinfo {pages} {125031} (\bibinfo {year} {2016})}\BibitemShut {NoStop}%
\bibitem [{\citenamefont {Huillery}\ \emph {et~al.}(2020)\citenamefont {Huillery}, \citenamefont {Delord}, \citenamefont {Nicolas}, \citenamefont {Van Den~Bossche}, \citenamefont {Perdriat},\ and\ \citenamefont {Hétet}}]{huillery_spin_2020}%
  \BibitemOpen
  \bibfield  {author} {\bibinfo {author} {\bibfnamefont {P.}~\bibnamefont {Huillery}}, \bibinfo {author} {\bibfnamefont {T.}~\bibnamefont {Delord}}, \bibinfo {author} {\bibfnamefont {L.}~\bibnamefont {Nicolas}}, \bibinfo {author} {\bibfnamefont {M.}~\bibnamefont {Van Den~Bossche}}, \bibinfo {author} {\bibfnamefont {M.}~\bibnamefont {Perdriat}},\ and\ \bibinfo {author} {\bibfnamefont {G.}~\bibnamefont {Hétet}},\ }\bibfield  {title} {\bibinfo {title} {Spin mechanics with levitating ferromagnetic particles},\ }\href {https://doi.org/10.1103/PhysRevB.101.134415} {\bibfield  {journal} {\bibinfo  {journal} {Physical Review B}\ }\textbf {\bibinfo {volume} {101}},\ \bibinfo {pages} {134415} (\bibinfo {year} {2020})}\BibitemShut {NoStop}%
\bibitem [{\citenamefont {Schmidt}\ \emph {et~al.}(2024)\citenamefont {Schmidt}, \citenamefont {Claessen}, \citenamefont {Higgins}, \citenamefont {Hofer}, \citenamefont {Hansen}, \citenamefont {Asenbaum}, \citenamefont {Uhl}, \citenamefont {Kleiner}, \citenamefont {Gross}, \citenamefont {Huebl}, \citenamefont {Trupke},\ and\ \citenamefont {Aspelmeyer}}]{schmidt_remote_2024}%
  \BibitemOpen
  \bibfield  {author} {\bibinfo {author} {\bibfnamefont {P.}~\bibnamefont {Schmidt}}, \bibinfo {author} {\bibfnamefont {R.}~\bibnamefont {Claessen}}, \bibinfo {author} {\bibfnamefont {G.}~\bibnamefont {Higgins}}, \bibinfo {author} {\bibfnamefont {J.}~\bibnamefont {Hofer}}, \bibinfo {author} {\bibfnamefont {J.~J.}\ \bibnamefont {Hansen}}, \bibinfo {author} {\bibfnamefont {P.}~\bibnamefont {Asenbaum}}, \bibinfo {author} {\bibfnamefont {K.}~\bibnamefont {Uhl}}, \bibinfo {author} {\bibfnamefont {R.}~\bibnamefont {Kleiner}}, \bibinfo {author} {\bibfnamefont {R.}~\bibnamefont {Gross}}, \bibinfo {author} {\bibfnamefont {H.}~\bibnamefont {Huebl}}, \bibinfo {author} {\bibfnamefont {M.}~\bibnamefont {Trupke}},\ and\ \bibinfo {author} {\bibfnamefont {M.}~\bibnamefont {Aspelmeyer}},\ }\href {https://doi.org/10.48550/ARXIV.2401.08854} {\bibinfo {title} {Remote sensing of a levitated superconductor with a flux-tunable microwave cavity}} (\bibinfo {year} {2024}),\ \bibinfo {note} {version Number: 2}\BibitemShut {NoStop}%
\end{thebibliography}%


\begin{thebibliography}{1}%
\makeatletter
\providecommand \@ifxundefined [1]{%
 \@ifx{#1\undefined}
}%
\providecommand \@ifnum [1]{%
 \ifnum #1\expandafter \@firstoftwo
 \else \expandafter \@secondoftwo
 \fi
}%
\providecommand \@ifx [1]{%
 \ifx #1\expandafter \@firstoftwo
 \else \expandafter \@secondoftwo
 \fi
}%
\providecommand \natexlab [1]{#1}%
\providecommand \enquote  [1]{``#1''}%
\providecommand \bibnamefont  [1]{#1}%
\providecommand \bibfnamefont [1]{#1}%
\providecommand \citenamefont [1]{#1}%
\providecommand \href@noop [0]{\@secondoftwo}%
\providecommand \href [0]{\begingroup \@sanitize@url \@href}%
\providecommand \@href[1]{\@@startlink{#1}\@@href}%
\providecommand \@@href[1]{\endgroup#1\@@endlink}%
\providecommand \@sanitize@url [0]{\catcode `\\12\catcode `\$12\catcode `\&12\catcode `\#12\catcode `\^12\catcode `\_12\catcode `\%12\relax}%
\providecommand \@@startlink[1]{}%
\providecommand \@@endlink[0]{}%
\providecommand \url  [0]{\begingroup\@sanitize@url \@url }%
\providecommand \@url [1]{\endgroup\@href {#1}{\urlprefix }}%
\providecommand \urlprefix  [0]{URL }%
\providecommand \Eprint [0]{\href }%
\providecommand \doibase [0]{https://doi.org/}%
\providecommand \selectlanguage [0]{\@gobble}%
\providecommand \bibinfo  [0]{\@secondoftwo}%
\providecommand \bibfield  [0]{\@secondoftwo}%
\providecommand \translation [1]{[#1]}%
\providecommand \BibitemOpen [0]{}%
\providecommand \bibitemStop [0]{}%
\providecommand \bibitemNoStop [0]{.\EOS\space}%
\providecommand \EOS [0]{\spacefactor3000\relax}%
\providecommand \BibitemShut  [1]{\csname bibitem#1\endcsname}%
\let\auto@bib@innerbib\@empty
\bibitem [{\citenamefont {Perdriat}\ \emph {et~al.}(2023)\citenamefont {Perdriat}, \citenamefont {Pellet-Mary}, \citenamefont {Copie},\ and\ \citenamefont {Hétet}}]{perdriat_planar_2023}%
  \BibitemOpen
  \bibfield  {author} {\bibinfo {author} {\bibfnamefont {M.}~\bibnamefont {Perdriat}}, \bibinfo {author} {\bibfnamefont {C.}~\bibnamefont {Pellet-Mary}}, \bibinfo {author} {\bibfnamefont {T.}~\bibnamefont {Copie}},\ and\ \bibinfo {author} {\bibfnamefont {G.}~\bibnamefont {Hétet}},\ }\bibfield  {title} {\bibinfo {title} {Planar magnetic {Paul} traps for ferromagnetic particles},\ }\href {https://doi.org/10.1103/PhysRevResearch.5.L032045} {\bibfield  {journal} {\bibinfo  {journal} {Physical Review Research}\ }\textbf {\bibinfo {volume} {5}},\ \bibinfo {pages} {L032045} (\bibinfo {year} {2023})}\BibitemShut {NoStop}%
\end{thebibliography}%

\end{document}


\title{Characterisation of a levitated sub-mg ferromagnetic cube \\in a planar alternating-current magnetic Paul trap: supplementary material}

\date{August 2024}

\author{Martijn Janse}
\author{Eli van der Bent}
\author{Mart Laurman}
\author{Robert Smit}
\author{Bas Hensen} \email{hensen@physics.leidenuniv.nl}
\address{Leiden Institute of Physics, Leiden University, P.O. Box 9504, 2300 RA Leiden, The Netherlands}

\maketitle

\section{Spectral information of parameters sweeps}\label{SM_A}
\begin{figure}[h]
    \centering
    \includegraphics[scale=1.5]{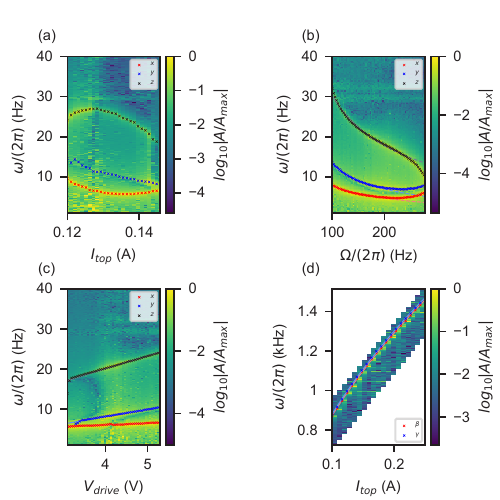}
    \caption{Dependency of eigenmode frequencies $\omega/(2\pi)$ on trap parameters. Here, we use the uncalibrated trap parameters, cf. the axes in Fig.~2 in the main text and Fig.~\ref{Figure3SM}. For the trap parameters of every measurement and for the fits, see the caption of Fig.~2 in the main text.} 
    \label{Figure1SM}
\end{figure}

In Fig.~\ref{Figure1SM}, we show the full spectral information of the measurements shown in Fig.~2 in the main text. The eigenmode frequency of every distinct mode is found by fitting a Lorentzian function to the spectrum in an initial specified spectral window and using this frequency to define a new window that is shifted in spectral range according to the expected scaling. For the librational modes we have only measured the spectrum partially to reduce the total measurement time.\\

\begin{figure}[h]
    \centering
    \includegraphics[scale=1.5]{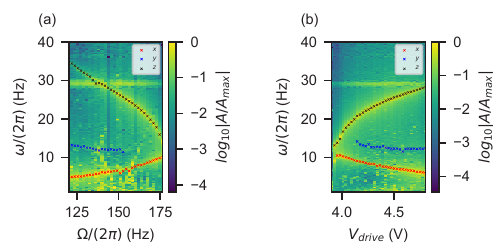}
    \caption{Dependency of eigenmode frequencies $\omega/(2\pi)$ on trap parameters at a different magnetic field gradient $B_{0}^{'} = 53$ mT/m and $B_{0} = 5.3$ mT. Here, we use the uncalibrated trap parameters, cf. the axes in Fig.~2 in the main text and Fig.~\ref{Figure3SM}. Both measurements are performed at current ratio $\xi = 2.2 \pm 0.05$. For \textbf{(a)}, $I_\text{trap} = 0.97$ A (or $V_\text{drive} = 4.5$ V) and for \textbf{(b)}, $\Omega/(2\pi) = 150$ Hz.}
    \label{Figure2SM}
\end{figure}

In Fig.~\ref{Figure2SM}, the spectra as a function of the trap frequency $\Omega$ and trap voltage $V_\text{drive}$ are given for a different magnetic field gradient $B_{0}^{'} = 53$ mT/m. Note how the behavior of the magnet is different from that in Fig.~\ref{Figure1SM}. The range of both $\Omega$ and $V_\text{drive}$ over which the magnet could be stably levitated, respectively $\Omega/(2\pi) = \{120,175\}$ Hz and $V_\text{drive} = \{3.9,4.8\}$ V, is smaller for $B_{0}^{'} = 53$ mT/m in comparison to $B_{0}^{'} = 81$ mT/m. Moreover, in both cases the scaling seems to not follow the scaling expected from Eqns.~1 and 2 in the main text and seems to be dominated by the converging-mode behavior we also started to see in Fig.~\ref{Figure1SM} for $\Omega/(2\pi) > 200$ Hz.

\section{Calibration of $B_{0}$, $B_{0}^{'}$ and $|B_{1}^{''}|$}\label{SM_B}
\begin{figure}[h]
    \centering
    \includegraphics[scale=1.5]{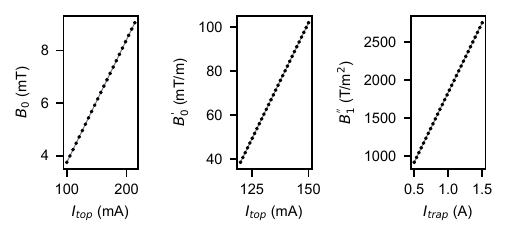}
    \caption{Conversion of the current $I_\text{top}$ to \textbf{(a)} magnetic field $B_{0}$ for $I_\text{bottom} = I_\text{top} - 0.040$ mA (constant magnetic field gradient), \textbf{(b)} magnetic field gradient $B_{0}^{'}$ for constant $I_\text{bottom} = 0.100$ mA and \textbf{(c)} magnetic field curvature $|B_1^{''}|$ for constant current ratio $\xi = 2.2$. The calibrations for \textbf{(a)-(b)} are based on a COMSOL Multiphysics model of the top and down coils in Helmholtz configuration. The calibration for \textbf{(c)} is based on a COMSOL Multiphysics model of the planar AC MPT.}
    \label{Figure3SM}
\end{figure}

In order to convert the measured parameters to the physical parameters that are used in Eqns.~1 and 2 in the main text, we have constructed a COMSOL Multiphysics model based on the exact dimensions of the MPT PCB and coils in Helmholtz configuration surrounding the MPT, shown in Fig.~\ref{Figure3SM}.\\

One part of the model calculates the magnetic field in the z-direction $B_0$ and its gradient $B_0^{'}$ at $\vec{r} = 0$ (center of the MPT) generated by the coils in Helmholtz configuration. The values of $B_0$ have been crosschecked by measurements with a Gauss meter. Another part of the model, that is also used for panel (b) of Fig.~1 in the main text, calculates the magnetic field curvature in the z-direction $|B_{1}^{''}|$ at $\vec{r} = 0$ generated by the trap loops. In both models, the derivatives of the magnetic fields, along a cut line in the z-direction for $x = 0$ and $y = 0$, are taken by using a Python interface connecting to COMSOL Multiphysics.\\

The simulated values of $|B_1^{''}|$ can be compared to the value that follows from the calculation in the main text. The derivation using the $\omega_\text{z}$ mode gives $|B_1^{''}| = 1335$ T/m$^2$ at $I_\text{trap} = 1.07$ A, whereas the simulated value at this current is $|B_1^{''}| = 1968$ T/m$^2$. The measured value for the magnetic field curvature is thus within an order of magnitude of the simulated values. The simulation overestimates $|B_1^{''}|$ and thus the vibrational mode frequencies $\omega_\text{i}/(2\pi)$ by approximately a factor 2. This is also seen in the time-domain simulations in Fig.~\ref{Figure4SM}. The deviation may be explained by the fact that the magnet does not levitate exactly at $z = 0$, but lower or higher according to the applied magnetic field gradient by the coils in Helmholtz configuration.\\

\section{Time-domain simulations of ferromagnetic cube in a planar alternating-current magnetic Paul trap}
\label{SM_C}
\begin{figure}[h]
    \centering
    \includegraphics[width = \textwidth]{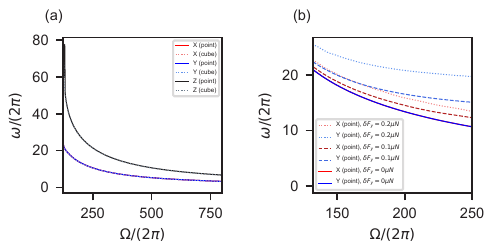}
    \caption{Time-domain simulations of the dependency of the vibrational eigenmodes $\omega/(2\pi)$ of the ferromagnet on the trap frequency $\Omega/(2\pi)$, as discussed for point particles in \cite{perdriat_planar_2023}. \textbf{(a)} Comparison of a point particle and a finite-volume cubic particle with an edge length of 250 $\mu$m. Due to the assumed symmetry in the simulation, the x and y modes overlap. \textbf{(b)} Effect of an offset force in the y-direction, $\delta F_\text{y}$, on the x and y mode eigenfrequencies. For $\delta F_\text{y} = 0$, the x and y modes overlap.}
    \label{Figure4SM}
\end{figure}

Following the equations of motion given in \cite{perdriat_planar_2023}, we have redone and improved upon the time-domain simulations in that work. The results are shown in Fig.~\ref{Figure4SM}. We have implemented a finite-volume version of the simulations for small angles $\beta$ and $\gamma$, in which we take the average magnetic force exerted on the ferromagnetic cube over the volume of the cube. We see that the results are very similar to the point-particle simulations. Moreover, we see that for large $\Omega$, the scaling of the eigenmodes does not change the expected $\Omega^{-1}$ according to Eqn.~1 in the main text, in contrast to the measurements. Therefore, the observed behavior cannot be explained by these time-domain simulations.\\

Additionally, we have measured the effect of an offset in the y-direction, $\delta F_\text{y}$, on the x and y mode eigenfrequencies to simulate the out-of-center position of the magnet in the MPT. We observe that for a finite $\delta F_\text{y}$ the eigenfrequencies of the x and y modes are different due to the asymmetry, as expected. More interestingly, we also see that for larger $\delta F_\text{y}$, the slope of the $\Omega^{-1}$ scaling decreases. This might explain the deviation of the slope ratio from the theoretically expected value $\omega_\text{x}:\omega_\text{y}:\omega_\text{z} = 1:1:2$ (Eqn.~1 in the main text), which is based on symmetry arguments. It can, however, not explain the observed change in scaling function, i.e. deviating from $\Omega^{-1}$ and $I_\text{trap}$ for too low $I_\text{trap}$ and too high $\Omega$.\\

\section*{REFERENCES}
\bibliography{01_bare-references}